\documentclass[11pt]{article}

\usepackage{fullpage}
\usepackage{setspace}
\usepackage{parskip}
\usepackage{titlesec}
\usepackage[section]{placeins}
\usepackage{xcolor}
\usepackage{breakcites}
\usepackage{lineno}
\usepackage{hyphenat}

\usepackage{times}

\PassOptionsToPackage{hyphens}{url}
\usepackage[colorlinks = true,
            linkcolor = blue,
            urlcolor  = blue,
            citecolor = blue,
            anchorcolor = blue]{hyperref}
\usepackage{etoolbox}
\makeatletter
\patchcmd\@combinedblfloats{\box\@outputbox}{\unvbox\@outputbox}{}{%
  \errmessage{\noexpand\@combinedblfloats could not be patched}%
}%
\makeatother

\usepackage{natbib}

\renewenvironment{abstract}
  {{\bfseries\noindent{\abstractname}\par\nobreak}\footnotesize}
  {\bigskip}

\titlespacing{\section}{0pt}{*3}{*1}
\titlespacing{\subsection}{0pt}{*2}{*0.5}
\titlespacing{\subsubsection}{0pt}{*1.5}{0pt}

\usepackage{authblk}

\usepackage{graphicx}
\usepackage[space]{grffile}
\usepackage{latexsym}
\usepackage{textcomp}
\usepackage{longtable}
\usepackage{tabulary}
\usepackage{booktabs,array,multirow}
\usepackage{amsfonts,amsmath,amssymb}
\providecommand\citet{\cite}
\providecommand\citep{\cite}

\newif\iflatexml\latexmlfalse

\AtBeginDocument{\DeclareGraphicsExtensions{.pdf,.PDF,.eps,.EPS,.png,.PNG,.tif,.TIF,.jpg,.JPG,.jpeg,.JPEG}}

\usepackage[utf8]{inputenc}
\usepackage[english]{babel}

\usepackage{float}

\usepackage[margin=1.5in]{geometry}

\usepackage{xcolor}
\iflatexml
\else
\fi

\begin{document}

\title{Combating Disinformation in A Social Media Age}

\author[1]{Kai Shu}%
\author[2]{Amrita Bhattacharjee}%
\author[2]{Faisal Alatawi}%
\author[3]{Tahora Nazer}%
\author[2]{Kaize Ding}%
\author[2]{Mansooreh Karami}%
\author[2]{Huan Liu}%
\affil[1]{Department of Computer Science, Illinois Institute of Technology, Chicago, Illinois, kshu@iit.edu}%
\affil[2]{Computer Science and Engineering, Arizona State University, Tempe, Arizona, {abhatt43, faalataw, kding9, mkarami, huanliu}@asu.edu}%
\affil[3]{Spotify, Boston, Massachusetts, tahoranazer@spotify.com}%

\vspace{-1em}

  \date{}

\begingroup
\let\center\flushleft
\let\endcenter\endflushleft
\maketitle
\endgroup

\selectlanguage{english}
\begin{abstract}
{The creation}, dissemination, and consumption of disinformation and
fabricated content on social media is a growing concern, especially with
the ease of access to such sources, and the lack of awareness of the
existence of such false information. In this paper, we present an
overview of the techniques explored to date for the combating of
disinformation with various forms. We introduce different forms of
disinformation, discuss factors related to the spread of disinformation,
elaborate on the inherent challenges in detecting disinformation, and
show some approaches to mitigating disinformation via education,
research, and collaboration. Looking ahead, we present some promising
future research directions on disinformation.%
\end{abstract}%

\sloppy

\section{Introduction}
%by Amrita
The proliferation and prevalence of social media in almost every facet of human lives have made the consumption of news and information extremely convenient to the users of such technology. The ease with which information, alerts, and warnings can be broadcasted to millions of people in a very short amount of time, has made social media a brilliant platform for information diffusion, especially for time-sensitive situations, for example, during natural disasters and crisis events. Given that a considerable fraction of individuals across the world use social media as a source of news \hyperref[csl:1]{(Shearer \& Matsa, 2018}; \hyperref[csl:2]{Nami Sumida \& Mitchell, 2019)}, and thus letting such news affect their opinions and actions, directly and indirectly, checking the veracity of such news becomes an important task. 

Over the last few years, the existence of disinformation online and the malicious agents acting as sources of such disinformation has been recognized and acknowledged. Research in the domain of disinformation detection and related fields has gained momentum, and different aspects of the problem are being approached by researchers from different perspectives. 

In this paper, we provide a comprehensive overview of the past and ongoing research in disinformation. We begin by defining a few relevant terms used extensively in the literature. In the following sections, we look at the history of disinformation and its characteristics in the social media age. Then we proceed to explain the challenges in the detection of disinformation, the different forms of disinformation that are prevalent on social media, \textcolor{black}{and detection and mitigation approaches and techniques}. We further talk about the factors behind the rapid spread of disinformation, steps taken to educate people about online disinformation, and conclude the review with some possible areas of future work.

\subsection{Definitions}

Upon a thorough review of existing literature in the context of the detection of deliberately fabricated content, we frequently come across the following keywords and terminologies - \textit{Misinformation}, \textit{Disinformation}, \textit{Fake News}, \textit{Hoax}, \textit{Rumor}, and \textit{Conspiracy theory}. In this section, we proceed to define these terms in the way these have been used by most researchers.

\textit{Misinformation} \hyperref[csl:3]{(Wu, Morstatter, Carley, \& Liu, 2019)} has been described in the literature as ``false, mistaken or misleading" information \hyperref[csl:4]{(Fetzer, 2004)}, often considered as an ``honest mistake."

On the other hand, \textit{disinformation} is false information, spread deliberately with the intention to mislead and/or deceive \hyperref[csl:5]{(Hernon, 1995)}. \textit{Fake news} has been defined as ``news articles that are intentionally and verifiably false, and could mislead readers'' \hyperref[csl:6]{(Allcott \& Gentzkow, 2017)}, and most researchers follow this definition. So fake news is an example of disinformation. In this paper, we use the terms ``disinformation'' and ``fake news'' interchangeably.

%Oxford English Dictionary defines Disinformation as ``False information which is intended to mislead, especially propaganda issued by a government organization to a rival power or the media'', while \textit{Fake News}, a term that gained popularity fairly recently, has been defined as ``False information that is broadcast or published as news for fraudulent or politically motivated purposes''.

Many researchers also use the term ``hoax'' (originating from \textit{hocus}, which means to trick or deceive), to refer to deliberate false information \hyperref[csl:7]{(Pratiwi, Asmara, \& Rahutomo, 2017}; \hyperref[csl:8]{Santoso, Yohansen, Nealson, Warnars, \& Hashimoto, 2017}; \hyperref[csl:9]{Vedova et al., 2018}; \hyperref[csl:10]{Kumar, West, \& Leskovec, 2016)}. Hoaxes are messages created with the intention of being spread to a large number of people, to ``persuade or manipulate other people to do or prevent pre-established actions, mostly by using a threat or deception'' \hyperref[csl:11]{(Vukovi{\'{c}}, Pripu{\v{z}}i{\'{c}}, \& Belani, 2009}; \hyperref[csl:12]{Hernandez, Hernandez, Sierra, \& Ribagorda, 2002)}. 

There has also been some work in the modeling of the spread of ``rumors." According to \hyperref[csl:13]{(Rosnow, 1991)}, ``rumors'' are ``public communications that are infused with private hypotheses about how the world works."

Although we do not cover it in much detail, ``conspiracy'' and ``conspiracy theory'' are also terminologies that researchers have used in related works. According to \hyperref[csl:14]{(van der Tempel \& Alcock, 2015)}, conspiracy theories are beliefs that are largely disregarded by society. A conspiracy theory ``involves not only the denial of official, commonly-held explanations about the causes of an event, but also the attribution of the event to a plan devised by a group of agents with hidden, unlawful, and malevolent intent''. For details on how conspiracy theories spread and government responses so far, we direct readers to \hyperref[csl:15]{(Sunstein \& Vermeule, 2009)}.

\subsection{A Brief History of Disinformation}
%by Amrita
The existence of disinformation and fake news is not new \hyperref[csl:16]{(Taylor, 2016)}, and the term has been widely in use since 1950s \hyperref[csl:17]{(Manning, Manning, \& Romerstein, 2004)}. For several decades individuals, groups and governments have tried to tarnish public opinion by exposing them to falsified, forged information as a way to sway people's political alignment. Deception and disinformation have been used by political forces for decades to have an edge over opponents/enemies - one of the famous documented instances of deception that succeeded in its mission was Operation Bodyguard, planned and executed by the Allied forces during World War II. The years following World War II, including the Cold War era, saw frequent use of forged, falsified information, clandestine radio stations, and newspaper articles to mislead the public and also governments. Several instances of such deceit and forgery carried out by governments have been documented in reports published by the CIA and the United States Bureau of Public Affairs \hyperref[csl:18]{({Department of State.}, 1981)}. Another infamous historical disinformation campaign was the Operation INFEKTION, active in the 1980s, which misled the public to believe that HIV/AIDS was developed by the US as a biological weapon \hyperref[csl:19]{(Boghardt, 2009)}.

Instances of political incidents where disinformation was prevalent, thus affecting public sentiment and opinions, include the assassination of president John F. Kennedy. Fetzer \hyperref[csl:4]{(Fetzer, 2004)} explains his work on this issue and the kinds of disinformation he encountered in documents released on the president's death. He further talks about studies where a movie of the assassination has been questioned, and claims have been made that the movie has been doctored - several frames have been edited out and the sequence of frames has been changed.

\subsection{Disinformation in the Age of Social Media}
%by Amrita
While the advent of the internet came as a boon to society, and its gradual adoption has resulted in a more connected world, the reachability of the internet also means it could be misused successfully. The internet provides a platform for fast dissemination of information and has often been referred to as the ``information highway'' in literature \hyperref[csl:20]{(De Maeyer, 1997)}. The positive impact of this is profound - with individuals becoming more aware of local and global news, their rights, raising awareness regarding global concerns including climate change and plastic pollution - which ultimately resulted in movements and aimed at taking action. However, all of these are based on the unifying assumption that the information available to people is real and not designed to mislead. 

The connectivity of the internet has been misused as well, to spread propaganda, to fulfill some individual or group agenda. There have been countless incidents of human lives being at stake due to false sensationalized information being spread via social media.
Hoax messages and rumors on messaging platforms like WhatsApp and Facebook spread like wildfire, and with an increase in smartphone usage, this makes the gullible public treat these false claims as genuine information and accordingly form opinions and take actions. Recently there have been several reported instances of vigilante justice where innocent people were lynched to death by an infuriated mob after rumors about the victims committing some kind of crime went viral. Social media platforms like WhatsApp and Facebook have been targeted by miscreants to spread rumors and hoaxes \hyperref[csl:21]{(Parth M.N., 30AD)}. 
%The most affected are the uneducated section of society who are unable to distinguish between genuine and falsified news that is disseminated via these platforms.
Some brutal incidents include one where a 21-year old innocent boy in a Mexican village was beaten and burned to death by an enraged mob \hyperref[csl:22]{(Patrick J. McDonnell, 20AD)}. Similar cases of lynchings and mob violence based solely on false claims and hoaxes propagated via WhatsApp have been a matter of concern in India, which happens to be the largest market for the company, having about 200 million active monthly users. Since this has become a growing concern, many studies have been conducted with a focus on how to detect and thwart the spread of disinformation and hoaxes \hyperref[csl:23]{(Arun, 2019)}, particularly on social media \hyperref[csl:24]{(Shu, Sliva, Wang, Tang, \& Liu, 2017}; \hyperref[csl:25]{Yang et al., 2019}; \hyperref[csl:26]{Shu \& Liu, 2019)}, but arriving at a unified solution has been challenging. 

\subsection{The Challenges of Detecting Disinformation }
We can categorize the challenges of detecting disinformation into two categories 1) content-related and 2) user-related challenges. The content of disinformation, in many cases, is highly sensationalized and is written using extreme sentiments, usually to affect the reader, which makes them interact with the post more \hyperref[csl:24]{(Shu, Sliva, Wang, Tang, \& Liu, 2017)}. Thus, such posts containing fabricated content often become ``viral'' and ``trending'' on social media \hyperref[csl:27]{(Vosoughi, Roy, \& Aral, 2018)}. In addition to that, the low cost of creating disinformation sources and the ease of using software-controlled social media bots to help spread disinformation \hyperref[csl:28]{(Shao, Ciampaglia, Varol, Flammini, \& Menczer, 2017)}.  From the user perspective, social media users are susceptible to disinformation, and they often lack awareness of disinformation \hyperref[csl:29]{(Sharma et al., 2019)}.
 
Disinformation such as fake news tends to be much more novel than the truth across all novelty metrics \hyperref[csl:27]{(Vosoughi, Roy, \& Aral, 2018)}.  People like to share novel news because new information is more valuable to people from an information-theoretic perspective.  Furthermore, novel news is valuable from a social perspective where people like to project the image of a person who is ``in the know,'' giving them a unique social status \hyperref[csl:27]{(Vosoughi, Roy, \& Aral, 2018)}.  People tend to trust their social contacts, and ``information'' that appear to be trendy. Therefore, spreading disinformation needs convincing content that looks like real articles and a method to make this news go viral \hyperref[csl:28]{(Shao, Ciampaglia, Varol, Flammini, \& Menczer, 2017)}. 

The goal of the early detection of disinformation is to prevent its further propagation on social media by giving early alerts of disinformation during the dissemination process \hyperref[csl:26]{(Shu \& Liu, 2019)}. Early detection of disinformation is extremely important to minimize the number of people influenced and hence minimize the damage. This becomes a challenge for automatic detection of disinformation as such detectors do not have access to user reactions, responses or knowledge bases for fact-checking early on, which could have been helpful for detection. However, exposing people to disinformation might help to increase the true positive rate of the disinformation detectors \hyperref[csl:30]{(Kim, Tabibian, Oh, Sch{\"o}lkopf, \& Gomez-Rodriguez, 2018)}. 
\color{black}
Furthermore, models such as epidemiological models of disease spread \hyperref[csl:31]{(Jin, Dougherty, Saraf, Cao, \& Ramakrishnan, 2013}; \hyperref[csl:32]{Bettencourt, Cintr{\'o}n-Arias, Kaiser, \& Castillo-Ch{\'a}vez, 2006)} and diffusion models like the independent cascade model \hyperref[csl:33]{(Goldenberg, Libai, \& Muller, 2001)} \hyperref[csl:34]{(Goldenberg, Libai, \& Muller, 2001)} and the linear threshold model \hyperref[csl:35]{(Granovetter, 1978}; \hyperref[csl:36]{Zafarani, Abbasi, \& Liu, n.d.)}, representing the propagation and spread of disinformation have also been explored to understand and potentially mitigate the spread of disinformation \hyperref[csl:37]{(Nguyen, Yan, Thai, \& Eidenbenz, 2012)}. However, these models can only be adequately studied after a significant amount of disinformation spread has already occurred, and therefore, this approach of detecting fake news/disinformation is not the most viable.
\color{black}
Another challenge behind fake news detection is that some keywords and ways of expression are specific to a certain kind of event or topic. So when a fake news classifier is trained on fake vs. real articles based on a particular event or topic, the classifier ends up learning the event-specific features and thus may not perform well when applied to classify fake vs. real articles corresponding to another kind of event. So, in general, fake news classifiers need to be generalized to be event-independent.

\color{black}

An important aspect of social media platforms that needs to be considered is the existence of filter bubbles or echo chambers created as a result of recommender systems on these platforms. Given the abundance of different kinds of content available online, social media platforms aim to use algorithms that would let users view and interact with content that is most relevant to them. This usually means that users would be exposed to posts, articles and viewpoints that align with their own beliefs, thus reinforcing them in the process, while also remaining unaware of opposing narratives and beliefs \hyperref[csl:38]{(Pariser, 2011)}. These `echo-chambers' or `filter bubbles' make the task of disinformation detection and mitigation especially difficult. Being exposed to the same viewpoint repeatedly would only reinforce the pre-existing beliefs of readers, and they would resist changing their opinion even if the narrative is later proven to be false by some fact-checking organization.

\color{black}

\section{Disinformation in Different Forms}
%by Amrita
Disinformation can exist in different forms - false text, images, videos, etc. In addition, there are different ways by which false content can be generated - both by humans and machines or a combination of the two. This makes the detection process more complex because most detection mechanisms assume some model of generation. In the following sections, we describe the most prevalent forms of false content on social media, and some detection techniques that have been explored by researchers.

\subsection{GAN Generated Fake Images}

Creating fabricated images and editing images into morphed representations can entirely change the semantic meaning of the image, and such doctored images are often propagated on social media. These images can be created by human agents or machine-generated e.g., by GANs (Generative Adversarial Networks) \hyperref[csl:39]{(Goodfellow et al., 2014)}. If such computer-generated images are realistic enough, it becomes very difficult for a human viewer to distinguish between real and such artificially generated images. It has been established that detecting GAN generated fake images is more challenging than images manually modified by human agents \hyperref[csl:40]{(Tariq, Lee, Kim, Shin, \& Woo, 2018}; \hyperref[csl:41]{Li, Chang, \& Lyu, 2018)}. This is because GANs generate a single image as a whole, so conventional techniques of detecting whether an image has been hand-edited or not, like analyzing metadata of the image, analyzing modifications on the frequency domain, finding disparities in JPEG quality of different regions of the same image, etc. fail when it comes to images generated by GANs. There is considerable ongoing work on the detection of GAN generated images, and we summarize a few of the promising approaches.

Marra et al. \hyperref[csl:42]{(Marra, Gragnaniello, Cozzolino, \& Verdoliva, 2018)} have discussed and compared the performance of several fake image detectors on a dataset of images comprising of both real and computer-generated images. The ``fake'' images were generated using a CycleGAN, described in \hyperref[csl:43]{(Zhu, Park, Isola, \& Efros, 2017)}, where an image from one domain was translated into an image from a target domain, and a similar inverse translation was done on the transformed image to convert it back to the original image. 
%This involves learning two mapping functions $G$ and $F$ (i.e., two generators to map from one domain to another) and two discriminators that learn to distinguish between generated samples and the actual data distribution. 
Because images uploaded on most social networking sites undergo some form of compression, Marra et al. \hyperref[csl:42]{(Marra, Gragnaniello, Cozzolino, \& Verdoliva, 2018)} conducted experiments comparing the fake image detectors on the compressed version of images as well and compared their performance. It was seen that, in case of compressed images, even with an overall decline in the performance of all the detectors studied, XceptionNet \hyperref[csl:44]{(Chollet, 2017)} gave quite accurate results.

While this work provides valuable insight into the efficacy of several image detectors on artificially generated images and images that have undergone compression, it has assumed the use of one particular GAN model for image generation. However, in real-world fake image detection, it is almost impossible to get access to the model used by the attacker, or even guess what the model could have been. Without this prior knowledge, it becomes very difficult to train classifiers to differentiate between fake and real images that would perform well. Zhang et al. \hyperref[csl:45]{(Zhang, Karaman, \& Chang, 2019)} have extended concepts from Odena et al. \hyperref[csl:46]{(Odena, Dumoulin, \& Olah, 2016)}, which says that GAN generated images have ``checkerboard artifacts'' as a result of the upsampling(deconvolution) layers, and the presence of these artifacts can guide fake image detectors to distinguish between real and fake images.
%Odena et al. explained the formation of spatial checkerboard patterns in GAN generated images, while Zhang et al. extended this to show that the frequency spectrum of such images also possesses significant artifacts. 
Utilizing these artifacts, the authors in \hyperref[csl:45]{(Zhang, Karaman, \& Chang, 2019)} have conducted experiments to classify real vs. fake GAN generated images, achieving varying degrees of success for images from different domains.

Other related approaches attempted to address this problem of detection of GAN generated images include the use of co-occurrence matrices by Nataraj et al. \hyperref[csl:47]{(Nataraj et al., 2019)}. Co-occurrence matrices were computed on the image pixels for each of the three color channels with the assumtion that co-occurrence matrices for GAN generated fake images would differ significantly from those of original images, and this would enable a fake image detector to learn features from the matrices and hence differentiate between real and fake images. Tariq et al. \hyperref[csl:40]{(Tariq, Lee, Kim, Shin, \& Woo, 2018)} have used an ensemble of shallow CNN based classifiers to detect ``real'' vs. ``fake'' faces generated by both GANs and by humans using professional image processing software.

\subsection{Fake Videos and Deepfakes}

Recent advancements in computer vision have enabled generative adversarial networks to learn to replace faces in videos with that of someone else, and ultimately create realistic videos that can easily fool gullible viewers into thinking it is genuine. This technology, also referred to as Deepfake, has been mostly used for malicious purposes - from using celebrities' faces in pornographic videos \hyperref[csl:48]{(Lee, 3AD)}, to creating fake videos of politicians, which has the potential to shift public sentiment and affect election procedures \hyperref[csl:49]{(George, 13AD)}. The advances in the deepfake video generation are being complemented by equally active research in the detection of such artificially generated videos. Here, we summarize a few of the recent notable works.

Deepfake videos generated using GANs often have certain imperfections if not post-processed correctly. For videos with human subjects involved in some kind of activity, the face swap by the GAN is done on a per-frame basis. So the dependency between consecutive frames is not utilized, and hence, can result in discrepancies between frames, for example, in the hue and/or illumination of the face region. 
%Although in most cases, this imperfection may be indiscernible to a human viewer of the final deepfake video, Convolutional Neural Networks or CNNs as feature extractors may be able to point out these discrepancies. 
The idea of across-frame inconsistencies has been used by G\"{u}era and Delp \hyperref[csl:50]{(G{\"u}era \& Delp, 2018)}. The authors used a ``convolutional LSTM'' consisting of a CNN to extract framewise features, which are then used as sequential input to an LSTM, and outputs from the LSTM are used to detect whether the video is a deepfake video or a non-manipulated one. 

Li et al. \hyperref[csl:41]{(Li, Chang, \& Lyu, 2018)} have attempted to detect fake videos by detecting eye blinking, with the assumption that in an artificially generated video, eye blinking may not be present or may be somewhat abnormal. This is because the models that generate fake videos by replacing someone's face in an already existing video, usually do not have access to images where the face has eyes closed. 
%Given that blinking is periodic and there is a temporal dependence among consecutive frames in the video, the authors used Long term Recurrent Convolutional Networks (LRCNs) to learn the temporal dependencies between states of eyes in frames. This can then, in turn, be used to detect whether a video is real or has been generated by some deepfake generating model. 
Another approach based on detecting discrepancies between real and artificially generated videos involves the analysis of 3D head pose in face videos \hyperref[csl:51]{(Yang, Li, \& Lyu, 2019)}. ``Facial landmarks'' were detected, which are points that map to features on the face - such as eyebrows, nose, mouth (which constitute the central part of the face), and the boundary or edge of the face. Deepfake conversion of a video usually replaces the central part of a face with that of a different face. During this conversion, there might arise some differences in alignment, which is not natural, and these result in inconsistent head poses. These fine misalignments can be difficult for a human to discern but can be learned by a model to distinguish between real and deepfake videos.

\subsection{Multimodal Content}

Most fake news articles and posts that are propagated online are composed of multiple types of data combined together - for example, a fabricated image along with a text related to the image. This is usually done to make the story/article more believable for the user viewing it. Moreover, there might also be comments on the post where some users may have pointed out that the article seems fabricated and fake, thus in a way guiding other users to verify the credibility of the source. This \textit{multimodal} nature of fake news articles is often quite challenging to handle when it comes to detection using deep learning based methods, and neural network architectures need to be carefully designed in order to make use of all the data available and the relationships between them. 

% add bold part under challenges in detecting fake news
Given the challenge of developing an event-independent disinformation detection model, Wang et al. \hyperref[csl:52]{(Wang et al., 2018)} designed a neural network architecture that has an event discriminator that tries to classify each post/article into the event that it is based on. The idea is to maximize the discriminator loss so that the model learns event-invariant features. The authors have addressed the multimodality of data on social media and have used both text and associated images, extracted event-invariant features from these and used it to train the fake news detector. Their experiments showed that multimodal detectors performed better than single-modality ones, and learning event-invariant features greatly  improved performance. Furthermore, their model performed significantly better than several state-of-the-art for multimodal fake news detection.

A similar but simpler approach using both image and text in news articles was explored by Yang et al. \hyperref[csl:53]{(Yang et al., 2018)}. The authors worked on a fake news dataset containing news articles collected from online sources, and used CNNs to extract features for both text and corresponding image input. Their experimental results also strengthen the idea that incorporating both image and text, and utilizing the multimodal nature of news, provides more information to fake news detectors and can improve performance.

Incorporating information from comments and reactions on posts can potentially improve the performance of fake news detectors. Sentiment analysis of these user comments can provide useful features to train classification models. A recent work by Cui et al. \hyperref[csl:54]{(Cui \& Lee, 2019)} is based on this idea - the authors have designed a model that takes into consideration user sentiments and similarity between articles based on user sentiments, along with features extracted from the image and text in the article. A modality discriminator, similar to the event discriminator in \hyperref[csl:52]{(Wang et al., 2018)}, has been used here to drive the modality distributions for image and text to get closer to each other. Their results showed that incorporating information from user sentiment did improve the fake news detector performance.

\section{Factors behind the Spread of Disinformation}
% whole of tahora's part

Social media users have a deficiency in spotting falsehood in specific emotional states, and when encountering what is consistent with their values or beliefs \hyperref[csl:55]{(Scheufele \& Krause, 2019)}. Malicious actors use this observation and target users with the same piece of disinformation on numerous occasions. Users who receive the same content from multiple sources have higher probability of believing and spreading it \hyperref[csl:56]{(Hasher, Goldstein, \& Toppino, 1977)}. This effective method can be further strengthened using social media bots. In this section we first focus on emotional factors that make social media users more vulnerable to disinformation and then discuss the role of bots in boosting the effect of disinformation and methods to prevent it. 

\color{black}
\subsection{Sources and Publishers}
Given the low cost of creating and publishing content online and the vast reach of social media platforms, several alternative media sources have emerged recently, often spreading false and/or highly biased claims. Although a large section of mainstream media is also politically/ideologically aligned towards either end of the political spectrum, these channels, having served as long-standing sources of information, do not intentionally publish false claims and disinformation. On the other hand, `alternative media' \hyperref[csl:57]{(Starbird, 2017)} has seen a rise in popularity and such media sources often publish false articles, opinions disguised as facts, and even highly polarizing conspiracy theories and pseudo-science related articles. These alt-media sources publish politically motivated articles, often directly challenging the narratives as published by mainstream media. Apart from the the intended target audience based on political ideology, these information sources often receive viewership from consumers of mainstream media, thus potentially fueling the reader's distrust in legitimate media sources \hyperref[csl:58]{(Haller \& Holt, 2019)}. In this context, assessing the credibility and trustworthiness of information sources is of paramount importance. There has been some effort among research communities to determine the political bias of news articles and media sources \hyperref[csl:59]{(Hirning, Chen, \& Shankar, 2017)} \hyperref[csl:60]{(Iyyer, Enns, Boyd-Graber, \& Resnik, 2014)}, and also on methods to assess credibility and trustworthiness of information online \hyperref[csl:61]{(Moturu \& Liu, 2009)} \hyperref[csl:62]{(Abbasi \& Liu, 2013)}. Apart from the ongoing research in academia, efforts are being put in by the journalism and fact-checking community \hyperref[csl:63]{(``{The Trust Project}'', n.d.)} \hyperref[csl:64]{(``{PolitiFact's fake news almanac - Infogram}'', n.d.)} \hyperref[csl:65]{(``{Media Bias/Fact Check - Search and Learn the Bias of News Media}'', n.d.)} to identify trustworthy sources and make readers aware of sources known to be publishing false articles, especially in the wake of the coronavirus pandemic.
%In the wake of the coronavirus pandemic, more of these alt-media sources have emerged, promoting conspiracy theories, pseudo-science based `cures' and 

\color{black}
\subsection{Emotional Factors}
Social media users have been widely affected by disinformation in recent years. For example, during the 2016 US presidential election, Guess et al. \hyperref[csl:66]{(Guess, Nyhan, \& Reifler, 2018)} observed that: 
\begin{enumerate}
    \item The average of 5.45 articles from fake news websites were consumed by the Americans age 18 or older.
    \item Many Americans visited fake news websites to complement their hard news, not to substitute them.
    \item There is a strong association between Facebook usage and fake news visits. 
    \item About half of the Americans who are exposed to fake news websites also visited a fact-checking websites.
\end{enumerate}

%\item Fake news websites were visited more likely by people whom supported Trump.
%\item Only one in four Americans read a fact-checking article in which the Trump supporters have less positive view in these kinds of outlets.
The question that we aim to understand is ``why do social media users believe disinformation?''. Familiarity is a driver of continued influence effects, making it a risk that repeating false information, even in a fact-checking context, may increase an individual's likelihood of accepting it as true \hyperref[csl:67]{(Swire, Ecker, \& Lewandowsky, 2017)}.
Based on an study by DiFonzo and Bordia~\hyperref[csl:68]{(DiFonzo \& Bordia, 2007)}, users are more prone to propagating disinformation when the situation is uncertain, they are emotionally overwhelmed and anxious, the topic of discussion is of personal importance to them, and they do not have primary control over the situation through their actions.

\textit{Uncertainty:} Spreading fake news can be a sense making activity in ambiguous situations and the frequency of fake news increases in uncertain situations, such as natural disasters or political events such as elections when people are unsure of the results \hyperref[csl:68]{(DiFonzo \& Bordia, 2007)}. When a crisis happens, people first seek information from official sources. However, in the lack of such information, they form unofficial social networks to make predictions with their own judgment and fill the information gap \hyperref[csl:69]{(Rosnow \& Fine, 1976)}. This might result in generating fake news such as a fake image of a shark swimming on the highways of Houston after Hurricane Harvey or millions of fake news posts that were shared on Facebook in the weeks leading to the US presidential election 2016. As the uncertainty increases, the reliance of firm beliefs and the unity among the users with the same ideology or in the same group reduces. Hence, users are more prone to accept new information, even false, as a compromise to resolve the uncertainty. Uncertainty can cause emotions such as anxiety and anger which will affect the spread of fake news in other ways \hyperref[csl:70]{(Marcus, 2017)}.

\textit{Anxiety:} Emotional pressure can play an important role in spreading fake news and can be triggered by emotions such as frustration, irritation, and anxiety. Anxiety can make people more prone to spreading unproved claims and less accurate in transmitting information \hyperref[csl:68]{(DiFonzo \& Bordia, 2007)}. In high anxiety situations, fake news can work as a justification process to relief emotional tension \hyperref[csl:71]{(Allport \& Postman, 1947)}. Fake news might be used as a method of expressing emotions in anxious situations that allows people to talk about their concerns and receive feedback informally; this process results in sense making and problem solving \hyperref[csl:72]{(Waddington, 2012)}. For example, during the devastating time of Hurricane Harvey, 2017, a fake news story accusing Black Lives Matter supporters of blocking first responders reaching the affected area was spread by more than one million Facebook users \hyperref[csl:73]{(Grenoble, 2017)}. Believing and spreading such fake news stories may help the people in disaster areas cope with the anxiety caused by delays in relief efforts \hyperref[csl:74]{(Fernandez, Alvarez, \& Nixon, 2017)}. The recent COVID-19 Coronavirus pandemic has also brought on a wave of disinformation, false claims and conspiracy theories to be propagated by anxious users on social media platforms, thus spreading more panic \hyperref[csl:75]{(Janosch Delcker \& Scott, 2020)}.
% We measure the anxiety of twitter users by LIWC Anxiety category.

\textit{Importance or outcome-relevance:} People pursue uncertainty reduction only in the areas that have personal relevance to them. For example, when a murder took place in a university campus, rumor transmission in the people from the same campus was twice the people who were from another university campus in the same city. Due to the difficulty of measuring importance, anxiety is often used as a proxy; being anxious about a fake news story shows importance \hyperref[csl:76]{(Anthony, 1973)}. 
% {\color{blue}Does fake news spread more near the location of the event under discussion?}

\textit{Lack of control:} Fake news represents ways of coping with uncertain and uncontrollable situations. When people do not have primary control over their situation (action-focused coping responses), they resort to secondary control strategies which are emotional responses such as predicting the worst to avoid disappointment and attributing events to chance. Two secondary control themes are explaining the meaning of events and predicting future events. 
% We use LIWC Future Focus category to measure the lack of control.

\textit{Belief:} Users prefer information that confirms their preexisting attitudes (selective exposure), view information consistent with their preexisting beliefs as more persuasive than dissonant information (confirmation bias), and are inclined to accept information that pleases them (desirability bias)~\hyperref[csl:77]{(Lazer et al., 2018)}. Moreover, building and maintaining social relations are vital to humans, hence, to ensure their reputation as a credible source of information, they tend to share the information in which they believe. Belief is found strongly related to transmission of rumors. When it comes to political issues people tend to rationalize what they \textit{want} to believe instead of attempting to find the truth. Hence, persuading themselves to believe what is aligned with their prior knowledge. This observation extends to the limit that appending corrections to misleading claims may worsen the situation; people who have believed the misleading claim may try to find reasons to dispute the corrections to an extent that they will believe in the misleading claim even more than before \hyperref[csl:78]{(Pennycook \& Rand, 2019)}.

% As for the misinformation in science they mention specific problematic areas causing misinformed citizens such as ``lack of knowledge about scientific facts", ``level of information on scientific process known as epistemic knowledge", ``inaccurate views or rejection of scientific consensus" and ``believing in conspiracy theories and endorsing them" among individuals.

\subsection{Bots on Social Media}
% from Faisal's part - edit the text to make it more `academic'

Thanks to the low cost of creating fake news sources and the software-controlled social media bots, it has never been easier to shape public opinion for political or financial reasons. Disinformation sources mimic mainstream media without obeying the same journalistic integrity \hyperref[csl:28]{(Shao, Ciampaglia, Varol, Flammini, \& Menczer, 2017)}. These sources rely on social bots to spread their content. %Low-credibility content has the same probability of going viral as fact-checked articles because they are not distinguishable. Although a big chunk of disinformation articles become viral, the vast majority of articles never have the chance to achieve their goal \hyperref[csl:28]{(Shao, Ciampaglia, Varol, Flammini, \& Menczer, 2017)}. 
The spread of fake news cannot be attributed to social bots only. However, curbing social bots is a promising strategy to slow down the propagation of disinformation. Actually, removing a small percentage of malicious bots can virtually eliminate the spread of low-credibility content \hyperref[csl:28]{(Shao, Ciampaglia, Varol, Flammini, \& Menczer, 2017)}.

Social bots mimic humans. They post content, interact with each other, as well as real people, and they target people that are more likely to believe disinformation. Therefore, people have a hard time distinguishing between the content posted by a human or a bot. The Botometer \hyperref[csl:28]{(Shao, Ciampaglia, Varol, Flammini, \& Menczer, 2017)}, formerly known as BotOrNot, is a machine learning tool that detects social bots on Twitter. 
%It gives each Twitter account a score between 0 and 1 that represents the level of automation. 
Bots use two strategies to spread low-credibility content; first, they amplify interactions with content as soon as it gets created to speed the process of making this article go viral. Second, bots target influential users in the hope of getting them to ``repost'' the fabricated article to increase public exposure and thus boosting its perceived credibility \hyperref[csl:28]{(Shao, Ciampaglia, Varol, Flammini, \& Menczer, 2017)}. 

% From Tahora's Part
In 2013, Twitter announced that about 5\% of its users are fake \hyperref[csl:79]{(Elder, 2013)}. The Wall Street Journal reported \hyperref[csl:80]{(Koh, 2014)} in March 2014 that half of the accounts created in 2014 were suspended by Twitter due to activities such as aggressive following and unfollowing behaviors which are known characteristics of bots \hyperref[csl:81]{(Lee, Eoff, \& Caverlee, 2011)}. In 2017, Varol et al. \hyperref[csl:82]{(Varol, Ferrara, Davis, Menczer, \& Flammini, 2017)} reported that between 9\% to 15\% of users on Twitter exhibit bot behaviors. In 2018,  Twitter suspended 70 million suspicious accounts \hyperref[csl:83]{(Timberg \& Dwoskin, 2018)} in an effort towards fighting fake news. 
% This swarm of bots rising on social media deviate the conversations and manipulate the activities of social media users.

Political disinformation is a major activity area for bots. 
%In a study of the 2010 US midterm elections on Twitter \hyperref[csl:84]{(Ratkiewicz et al., 2011)}, authors observed a group of bots smearing a candidates by promoting specific URLs. 
% These bots further targeted popular users who were active on the topic and mentioned them in their promotional tweets. When targeted users receive the same content from multiple sources the probability of involving in the cascade increases~\hyperref[csl:56]{(Hasher, Goldstein, \& Toppino, 1977)}. 
During the US presidential election in 2016, bots produced 1.7 billion tweets and successfully outnumbered the tweets supporting one candidate by the factor of four \hyperref[csl:85]{(Kelion \& Silva, 2016)}. Similarly on Facebook, millions of fake stories supporting each candidate were shared and the balance was 4 to 1 supporting the same candidate \hyperref[csl:6]{(Allcott \& Gentzkow, 2017)}. These findings raise the question whether fake news did \hyperref[csl:86]{(Parkinson, 2016)} or did not \hyperref[csl:6]{(Allcott \& Gentzkow, 2017)} affect the election results.

Bots, during the natural disasters of 2017, Mexico Earthquake and Hurricanes Harvey, Irma, and Maria, used disaster-related hashtags to promote political topics such as {\tt \#DACA} and {\tt \#BlackLivesMatter} \hyperref[csl:87]{(Khaund, Al-Khateeb, Tokdemir, \& Agarwal, 2018)}. These bots shared fake news and hoaxes such as a shark swimming in a flooded highway after Hurricanes Harvey and Irma. Bots accelerate the spread of rumors and increase the number users exposed to them by 26\% \hyperref[csl:27]{(Vosoughi, Roy, \& Aral, 2018)}. Bots also use misdirection and smoke screening to sway the public attention from specific topics \hyperref[csl:88]{(Abokhodair, Yoo, \& McDonald, 2015)}. During the Syrian civil war in 2012, bots tweeted about political and natural crisis happening worldwide while including keywords related to the Syrian civil war to sway attention from the civil war, i.e. misdirection. In smoke screening, bots talked about other events in Syria using relevant hashtags {\tt \#Syria} (in English and Arabic) but the content was not about the civil war. 

Activities of bots in spreading disinformation on social media has been widely reflected by news agencies and vastly studied by researchers. A list of major events infiltrated by bots and reported by news media is presented in Table \ref{tab:bot_news}. 

Researchers have studied bots on social media during numerous major events: natural disasters, man-made disasters such as civil wars and mass shootings, and political events such as Presidential Elections or referendums. 
% An extensive list of such research papers is presented in Table \ref{tab:bot_areas}.\selectlanguage{english}
\begin{table}%[t]
\caption{{News Articles on Bots on Social Media During Major Events.}}
\label{tab:bot_news}
\centering
%\makebox[\linewidth]{
\begin{tabular}{|| c | c | c ||}
     \toprule
     \bf{Outlet} & \bf{Category} & \bf{Event}\\\midrule
     The Guardian\footnote{https://www.theguardian.com/technology/2018/jan/19/twitter-admits-far-more-russian-bots-posted-on-election-than-it-had-disclosed} &Elections&US Presidential Election 2016\\
     % more than 50,000 Russia-linked accounts used its service to post automated material about the 2016 US election
     %https://www.theguardian.com/technology/2018/jan/19/twitter-admits-far-more-russian-bots-posted-on-election-than-it-had-disclosed
     Time Magazine\footnote{http://time.com/5286013/twitter-bots-donald-trump-votes/} & Elections & US Presidential Election 2016\\ % It is also on Brexit
     % bots added 1.76 percentage point to the pro-?leave? vote share as Britain weighed whether to remain in the European Union, and may explain 3.23 percentage points of the actual vote for Trump in the U.S. presidential race.
     Bloomberg\footnote{https://www.bloomberg.com/news/articles/2018-02-19/now-bots-are-trying-to-help-populists-win-italy-s-election}&Elections&Italian General Election 2018\\
     The Telegraph\footnote{https://www.telegraph.co.uk/technology/2018/10/17/russian-iranian-twitter-trolls-sent-10-million-tweets-fake-news/}&Referendum&Brexit\\
     Politico Magazine\footnote{https://www.politico.com/magazine/story/2018/02/04/trump-twitter-russians-release-the-memo-216935}&Propaganda&\#ReleaseTheMemo Movement 2018\\
     % https://www.politico.com/magazine/story/2018/02/04/trump-twitter-russians-release-the-memo-216935
     
     The Guardian\footnote{https://www.cnn.com/2018/10/19/tech/twitter-suspends-spam-khashoggi-accounts-intl/index.html}&Propaganda&Russia-Ukraine Conflict 2014\\
     CNN\footnote{https://www.cnn.com/2018/10/19/tech/twitter-suspends-spam-khashoggi-accounts-intl/index.html}&Propaganda&Jamal Khashoggi's Death 2018\\
     
     The Telegraph\footnote{https://www.telegraph.co.uk/business/2018/03/31/twitter-bots-manipulating-stock-markets-fake-news-spreads-finance/}&Fake News&Stock Market - FTSE 100 Index 2018\\

     Forbes\footnote{https://www.forbes.com/sites/thomasbrewster/2016/12/20/methbot-biggest-ad-fraud-busted/\#4e33e3004899}&Ad Fraud&AFK13 Attack\\

     The New York Times\footnote{https://www.nytimes.com/2018/02/19/technology/russian-bots-school-shooting.html}& Mass Shooting & Florida School Shooting 2018\\
     
     Fox News\footnote{https://www.foxnews.com/tech/fake-facebook-accounts-misinformation-spread-quickly-in-wake-of-santa-fe-school-shooting}&Mass Shooting & Texas School Shooting 2018\\
     %    
     % 12,000 automated Twitter accounts that are often used to spread disinformation, four of the top 10 phrases tweeted by bot or troll accounts in the immediate 24-hour aftermath were related to the Santa Fe shooting

     The Telegraph\footnote{https://www.telegraph.co.uk/news/2017/11/13/russian-bot-behind-false-claim-muslim-woman-ignored-victims/}&Terrorist Attack&Westminster Terror Attack 2017\\

     Medium\footnote{https://medium.com/hci-wvu/countering-fake-news-in-natural-disasters-using-bots-and-citizen-crowds-412bbef6b489}&Natural Disasters&Mexico City Earthquake 2017\\
     \bottomrule
    \end{tabular}
%     }

\end{table}

\subsubsection{Representative Bot Detection Methods}
Ferrara et al.~\hyperref[csl:100]{(Ferrara, Varol, Davis, Menczer, \& Flammini, 2016)} proposed a taxonomy of bot detection models which divides them into three classes: (1) graph-based, (2) crowdsourcing, and (3) feature-based social bot detection methods.

\noindent \textbf{Graph-based Methods} \\
Graph-based social bot detection models lie on the assumption that the connectivity of bots is different from human users on social media. \textit{SybilRank}~\hyperref[csl:101]{(Cao, Sirivianos, Yang, \& Pregueiro, 2012)} is a graph-based method proposed to efficiently detect adversary-owned bot accounts based on the links they form. The underlying assumption is that bots are mostly connected to other bots and have limited number of links to human users. 
%Hence, if we start short random walks from a set of trusted users in the network, there is a higher probability that we land on a human user rather than a bot. 
Bots also show different characteristics in the communities they form. In a study on the bots that were active during the natural disasters in 2017, Khaund et al.~\hyperref[csl:87]{(Khaund, Al-Khateeb, Tokdemir, \& Agarwal, 2018)} observed that bots form more hierarchical communities with cores of bots strongly connected to each other and peripheral members who are weakly connected to the core and to each other. Moreover, human users had more communities and their communities were more tightly knit.

\noindent \textbf{Crowdsourcing Methods} \\
Crowdsourcing social bot detection uses human annotators, expert and hire workers, to label social media users as human or bot~\hyperref[csl:102]{(Wang et al., 2013)}. This method is reliable and has near zero error when the inter annotator agreement is considered. However, it is time consuming, not cost effective, and not feasible considering millions of users on social media. Crowdsourcing and manual annotation are still being used as methods for collecting gold standard datasets for feature based bot detection models, most of which use supervised classification.

\noindent \textbf{Feature-based Methods} \\
Feature based social bot detection methods are based on the observation that bots have different characteristics than human users. 
% This difference is reported in various aspects: diffusion patterns based on retweet, mentions, and hashtag  co-occurrence~\hyperref[csl:103]{(Ratkiewicz et al., 2011)}, profile information such as language, geographic locations, and account creation time~\hyperref[csl:104]{(Thomas, Grier, \& Paxson, 2012)}, number of friends, followers, and their ratio~\hyperref[csl:81]{(Lee, Eoff, \& Caverlee, 2011)},  temporal patterns of content generation~\hyperref[csl:105]{(Lee \& Kim, 2014)}, content features such as linguistic aspects and part-of-speed tagging, latent topics, and sentiment of posts~\hyperref[csl:106]{(Morstatter, Wu, Nazer, Carley, \& Liu, 2016)}. These features have been widely used in feature based bot detection models that are based on supervised classification methods.
To use feature-based supervised bot detection models, one must identify differences among bot and human users in terms of features such as content or activity in a labeled dataset. Then, a classifier is trained on the features and labels to distinguish bots from humans in an unobserved dataset. Different classification methods can be used for this purpose such as Support Vector Machines~\hyperref[csl:106]{(Morstatter, Wu, Nazer, Carley, \& Liu, 2016)}, Random Forests~\hyperref[csl:81]{(Lee, Eoff, \& Caverlee, 2011)}, and Neural Networks~\hyperref[csl:107]{(Kudugunta \& Ferrara, 2018)}. We describe some common user features below:
\begin{itemize}
\item Content: The measures in this category focus on the content shared by users. Words, phrases~\hyperref[csl:82]{(Varol, Ferrara, Davis, Menczer, \& Flammini, 2017)}, and topics~\hyperref[csl:106]{(Morstatter, Wu, Nazer, Carley, \& Liu, 2016)} of social media posts can be a strong indicator of bot activity. Also, bots are motivated to persuade real users into visiting external sites operated by their controller, hence, share more URLs in comparison to human users~\hyperref[csl:108]{(Chu, Gianvecchio, Wang, \& Jajodia, 2012}; \hyperref[csl:103]{Ratkiewicz et al., 2011}; \hyperref[csl:109]{Xie et al., 2008)}. Bots are observed to lack originality in their tweets and have large ratio of retweets/tweets~\hyperref[csl:84]{(Ratkiewicz et al., 2011)}. 

\item Activity Patterns: Bots tweet in a ``bursty'' nature~\hyperref[csl:108]{(Chu, Gianvecchio, Wang, \& Jajodia, 2012}; \hyperref[csl:105]{Lee \& Kim, 2014)}, publishing many tweets in a short time and being inactive for a longer period of time. Bots also tend to have very regular (e.g. tweeting every 10 minutes) or highly irregular (randomized lapse) tweeting patterns over time~\hyperref[csl:110]{(Zhang \& Paxson, 2011)}. 

\item Network Connections: Bots connect to a large number of users hoping to receive followers back but the majority of human users do not reciprocate. Hence, bots tend to follow more users than follow them back~\hyperref[csl:108]{(Chu, Gianvecchio, Wang, \& Jajodia, 2012)}.
\end{itemize}

\section{Detecting Disinformation}
\label{sec_detection}
In this section, we discuss methods to detect disinformation. There are three areas that we discuss regarding detecting disinformation: 1) the users that disinformation is targeting, 2) the content of the disinformation itself, and 3) the way disinformation spreads over the network. Each one of these areas provides vital features that could be used to detect and combat disinformation. An excellent solution to the spread of disinformation would employ most, if not all, of these areas. However, most existing works had limited success to combat disinformation by focusing on one particular area out of the ones mentioned above to build their solutions. In this section, we discuss each one of these areas in detail. Although most fake news or disinformation detection techniques are supervised, some semi-supervised \hyperref[csl:111]{(Guacho, Abdali, Shah, \& Papalexakis, 2018)} and unsupervised \hyperref[csl:112]{(Hosseinimotlagh \& Papalexakis, 2018)} techniques have also been developed. We believe that this section could help to understand the methods used to detect disinformation, and it could guide new detectors to use more than one of these areas.  

% =============================
\subsection{The Role of Individuals to Detect Disinformation} 
In this section, we introduce how various aspects of users who are engaged in the spreading of disinformation can be utilized to detect disinformation.
\subsubsection{Modeling User Interactions}
To combat fake news, some social media outlets turn to users to flag potential fake news articles. For instance, Facebook recently introduced tools that help users to flag articles that they deem as fake \hyperref[csl:113]{(Tschiatschek, Singla, Gomez Rodriguez, Merchant, \& Krause, 2018)}. These tools generate crowd signals that could be used to train classifiers to flag fake news. Many fake news detectors rely on users' responses because they hold valuable information about the credibility of the news \hyperref[csl:114]{(Qian, Gong, Sharma, \& Liu, 2018)}. Castillo et al. \hyperref[csl:115]{(Castillo, El-Haddad, Pfeffer, \& Stempeck, 2014)} pointed out that users' responses to news articles hold valuable information that helps to understand the properties of the content itself. Also, users' stances and sentiment could assist in detecting disinformation \hyperref[csl:114]{(Qian, Gong, Sharma, \& Liu, 2018)}.

Shu et al. \hyperref[csl:116]{(Shu, Zhou, Wang, Zafarani, \& Liu, 2019)} used social media's user profiles for fake news detection. They measured the sharing behavior of users that spread disinformation. Shu et al. analyzed the explicit and implicit profile features between these user groups to use those features in detecting disinformation on social media.

Tschiatschek et al. \hyperref[csl:113]{(Tschiatschek, Singla, Gomez Rodriguez, Merchant, \& Krause, 2018)} developed Detective, a Bayesian inference algorithm that detects fake news and simultaneously learns about users over time. Detective selects a subset of the news to send to an expert, such as a third-party fact-checking organization, to determine if the articles are fake or not. Over time the algorithm learns about users flagging behavior to avoid the effect of adversarial users. The experiments show that Detective is robust even if the majority of users are malicious.

Many users try to combat disinformation with the truth. Some users share links to fact checking websites that debunk false claims, or disinformation. The fake checking articles come from sites like Snopes.com, Politifact.com, or FactCheck.org.  To stimulate the users to spread fact-checked content to other users, \hyperref[csl:117]{(Vo \& Lee, 2018)} propose an article recommendation model. The model provides personalized fact-checking URLs to each user based on their interests to encourage them to engage in fighting disinformation. 
%The recommendation model aims to promote new interesting fact-checking articles to the user. They could use these articles in their messages, correct unverified claims or misinformation, and spread fact-checked information. 

\subsubsection{Using User Sentiments}
User sentiment about news articles may help to detect disinformation. In a study by Vosoughi et al. \hyperref[csl:27]{(Vosoughi, Roy, \& Aral, 2018)} show that disinformation triggers different feelings in users than real news. Disinformation sparks fear, disgust, and surprise that could be observed from user responses. On the other hand, real news stimulates anticipation, sadness, joy, and trust. Some of the existing works utilize this phenomena to detect disinformation. 

Pamungkas et al. \hyperref[csl:118]{(Pamungkas, Basile, \& Patti, 2019)} argue that when users face disinformation, in this case rumors, they take different stances about the article. Some users support the rumor, while others deny it.  To predict user stance, Pamungkas et al. use conversation-based and affective-based features
%, covering different facets of effect.  
They classify user response in these classes: support, deny, query, or comment to the rumor.  This work could be used as training features to a disinformation detector.

%To reduce the spread of fake news, detectors should be able to detect them as early as possible. Users' responses hold information about the stances and the sentiment of those users that could assist in detecting disinformation. Therefore, several disinformation detectors rely on users' responses.  However, waiting for users to respond to potential fake news articles defies the purpose of stopping fake news before it spreads. It will be too late to do something about it.  
As we discussed previously, early detection of fake news is challenging, because detectors do not have access to user responses. However, users' responses to the previous articles exist and could be used to generate responses to news articles that could enhance the detection of fake news. User responses towards previously propagated articles may hold rich information and latent user intelligence that existing works ignore \hyperref[csl:114]{(Qian, Gong, Sharma, \& Liu, 2018)}. Qian et al. \hyperref[csl:114]{(Qian, Gong, Sharma, \& Liu, 2018)}  developed a detector that tackles the early fake news detection problem using only the news article's text. They developed a generative model that can be used to generate responses to new articles in order to assist in fake news detection. %The generative model assists in the early detection of fake news articles by simulating user responses to new articles. It leverages collective user intelligence on why articles must be real or fake. 

% Qian et al. \hyperref[csl:114]{(Qian, Gong, Sharma, \& Liu, 2018)} developed a detector that tackles the early fake news detection problem using only the news article's text. The detector is a Two-Level Convolutional Neural Network with User Response Generator (TCNN-URG).  The TCNN is used to obtain semantic information from the text, while the URG uses user responses to existing articles to learns a generative model of their responses. This generative model can be used to generate responses to new articles in order to assist in fake news detection. TCNN-URG combines the power of both the discriminative model and the generative model to detect fake news as early as possible. The discriminative model detects fake news using the text of the article and features extracted from it.  The generative model assists in the early detection of fake news articles by simulating user responses to new articles. It leverages collective user intelligence on why articles must be real or fake. The TCNN-URG leverage the relationship between the article text content and user responses it invokes. Therefore, TCNN-URG provides better understanding of the news article text and its veracity. 

% =============================
\subsection{Leveraging the Content to Detect Disinformation}
In this section, we will illustrate how to leverage disinformation content in the detection task. Also, we will talk about AI-generated disinformation.  We believe that AI-generated disinformation will be a very important topic for research in the near future. 

As we mentioned previously, one of the challenges of detecting disinformation is detecting fake news regarding newly emerged events. Most of the existing detection strategies tend to learn event-specific features that can not be transferred to unseen events. Therefore, they perform poorly on this challenge \hyperref[csl:52]{(Wang et al., 2018)}. To overcome this challenge, Wang et al. \hyperref[csl:52]{(Wang et al., 2018)} propose Event Adversarial Neural Network (EANN) that can derive event-invariant features. The event-invariant features help the detection of fake news on newly arrived events. Also, EANN can learn transferable features for unseen events. 

%EANN has three components: multi-modal feature extractor, fake news detector, and an event discriminator. The role of the multi-modal feature extractor is to examine the post derive textual and visual features. Along with the multi-modal feature extractor, the fake news detector learns the discriminable representation to perform its primary task of detecting fake news. The event discriminator's role is to eliminate the event-specific features and only retain shared features between the different events.

Although most disinformation content is created manually, there is potential for AI-generated textual content in the near future. In recent years, advances in natural language understanding and natural language processing has achieved great results for challenging language tasks such as neural text summarization and machine translation. However, the same technologies used for these applications could be used to generate fake news by adversaries.  We are on the verge of the era of Neural Generated Disinformation because anyone can easily and cheaply collect and process vast amounts of data \hyperref[csl:119]{(Zellers et al., 2019)}. In addition to that, the recent developments in text generation \hyperref[csl:120]{(Jozefowicz, Vinyals, Schuster, Shazeer, \& Wu, 2016}; \hyperref[csl:121]{Radford, Narasimhan, Salimans, \& Sutskever, 2018}; \hyperref[csl:122]{Radford et al., 2019)} lead Zellers et al.  \hyperref[csl:119]{(Zellers et al., 2019)} to develop a model to generate fake news. 

Zellers et al.  \hyperref[csl:119]{(Zellers et al., 2019)} developed a generative model called Grover, which is a controllable text generation model. The goal of this model to act as a threat model to study the threat of AI-generated news. This approach is inspired by threat modeling commonly used in the field of computer security.  
%The main goal of modeling the threats is to analyze the space of potential threats and vulnerabilities in a system to develop robust defenses.
Grover can write a news article, given only a headline. 
For example, the user might input the title `Link Found Between Vaccines and Autism' as a headline for a new fake news article. Grover then generates the body of the article and rewrites the headline of the article in a more appropriate way. 
%The generated articles fooled qualified readers, and they found those articles more trustworthy than fake news written by humans. 
%The best current discriminators can identify AI-generated fake news with 73\% accuracy. 
Grover can identify news generated by itself with around 90\% accuracy. AI-Generated content could possibly lead to other threats. For instance, AI could be used by an adversary to generate comments to news articles or even the articles themselves \hyperref[csl:119]{(Zellers et al., 2019)}.

% This finding led Zellers et al.  \hyperref[csl:119]{(Zellers et al., 2019)} to conclude that ``the best models for generating neural disinformation are also the best models at detecting it.''

% AI-Generated content could possibly lead to other threats. For instance, "an adversary might generate comments or have entire dialogue agents, they might start with a human-written news article and modify a few sentences, and they might fabricate images or video. These threat models ought to be studied by researchers also so that we can create better defenses."  \hyperref[csl:119]{(Zellers et al., 2019)}

% Also, AI could be used to fabricate images or videos.
% Researchers should studied these threat models to create better defenses.

Media and journalism companies are coming up with smartphone apps and websites that provide very short summaries of news articles, to satisfy readers who want to consume as much information possible in a short time. In this regard, neural text summarization seems to be a promising way of automating the news article summarization task. With the gradual improvement of such summarization models, an interesting new challenge would be to perform fake news detection on such neural generated summaries instead of on the actual article. Esmaeilzadeh et al. \hyperref[csl:123]{(Esmaeilzadeh, Peh, \& Xu, 2019)} have performed fake news detection on summaries generated from news articles using abstractive summarization models, and compared the detection performance when using the entire news text and headline text as input. For the dataset and text summarization model used in \hyperref[csl:123]{(Esmaeilzadeh, Peh, \& Xu, 2019)}, the model trained on the neural generated summaries performed best when it came to detection accuracy, hinting that the summarization step acts as some form of a feature generator. 

\color{black}

Apart from the ML based methods discussed above, significant efforts have been put in by the fact-checking, data management and database community \hyperref[csl:124]{(Lakshmanan, Simpson, \& Thirumuruganathan, 2019)} toward disinformation detection, leveraging structured data and knowledge graphs \hyperref[csl:125]{(Ciampaglia et al., 2015)}. Automatic fact-checking using \textit{knowledge graphs} in order to check the veracity of claims is an active area of ongoing research and when done successfully can potentially replace the tedious task of manual fact-checking done by experts and journalists. Such techniques include path analysis \hyperref[csl:126]{(Shi \& Weninger, 2016)} and link prediction \hyperref[csl:127]{(Shi \& Weninger, 2016)} in knowledge graphs among others. 

Attempts at disinformation detection via checking the credibility of claims made online has been done via methods in the domain of \textit{Truth Discovery and Fusion}. Some of these methods use multiple online sources, both supporting and refuting, to check the veracity of claims and identify the truth \hyperref[csl:128]{(Jin, Cao, Zhang, \& Luo, 2016)} \hyperref[csl:129]{(Yin, Han, \& Philip, 2008)}. Given the evolving nature of online social media content along with the problem of early-detection of disinformation, Zhang et al. have proposed a dynamic truth discovery scheme \hyperref[csl:130]{(Zhang, Wang, \& Zhang, 2017)} to deal with noisy and incomplete data available online. 
%Fact-checking using information from multiple sources that either support or refute the claim being examined 

\color{black}
\subsection{Exploiting Networks for Detecting Disinformation}
% by Amrita
In this section, we illustrate how to detect disinformation with the signals extracted from different types of networks.
\subsubsection{Detection of Fake Content via Propagation Networks}

%Most of the methods discussed above take into consideration the linguistic features of text and unnatural pixel qualities and discrepancies in artificially generated images as the determining factors as to whether a news article or post is fake or real. However, Often

 Malicious agents, who intend to spread fake fabricated news on social media, want rapid dissemination of the post/article, in order to reach as many people as possible quickly. Often bots and bot-like accounts are used to propagate the news faster. Keeping this in mind, the user profiles of users sharing the news, and also the propagation pattern can provide clues to determine the veracity of the article. One relatively recent work that achieved promising results using this approach is that of Liu and Wu \hyperref[csl:131]{(Liu \& Wu, 2018)}, where the authors have used the concept of propagation path of news articles in the detection of fake news. In this work, propagation path of an article is a sequence of tuples, each tuple consisting of a timestamp $t$ and a vector characterizing the user who shared the article at that timestamp. The propagation paths for the news articles, are fed separately into RNN and CNN units, and after subsequent pooling and concatenation operations, a classifier is trained on the learned representations in order to differentiate between fake and real articles. A similar approach using propagation structures or propagation trees, and features related to the user, original message and ``reposts'', has been experimented with using false rumor vs. real data from Weibo \hyperref[csl:132]{(Wu, Yang, \& Zhu, 2015)}. 
 %A set of 23 features related to the user, original message and reposts were extracted for each propagation tree and classifiers were trained in order to label the original post as fake or real. 

\subsubsection{Advanced Graph Mining for Disinformation Detection}

% Social media services (e.g., Facebook, Youtube) have emerged as popular platforms for content sharing and information dissemination. 
% The rapid growth of social media also provides malicious users a new and convenient medium to spread disinformation for their noxious intentions. For example, during the 2016 US presidential election, a lot of fake news about presidential candidates is spread on various social platforms~\hyperref[csl:133]{(Jin et al., 2017)}: 115 pro-Trump fake stories that were shared on Facebook a total of 30 million times, and 41 pro-Clinton fake stories shared a total of 7.6 million times are observed in \hyperref[csl:6]{(Allcott \& Gentzkow, 2017)}. Such a huge amount of widely spread disinformation have greatly destroyed the public persona of
% candidates and misled the judgment of voters. Detecting disinformation on social networks has become a critical research task, in order to block the spread and refute such disinformation. 
Recently, much efforts have been devoted on graph mining-based disinformation detection due to their superior capabilities. Graph neural networks (GNNs), a family of neural models for learning latent node representations in a graph, have been widely used in different graph learning tasks and achieved remarkable success~\hyperref[csl:134]{(Kipf \& Welling, 2016}; \hyperref[csl:135]{Ding, Li, Bhanushali, \& Liu, 2019}; \hyperref[csl:136]{Ding, Li, Li, Liu, \& Liu, 2019}; \hyperref[csl:137]{Monti, Frasca, Eynard, Mannion, \& Bronstein, 2019}; \hyperref[csl:138]{Bian et al., 2020)}. Due to the superior modeling power, researchers have proposed to leverage GNNs for solving the disinformation detection problem. As one of the first endeavors, \hyperref[csl:137]{(Monti, Frasca, Eynard, Mannion, \& Bronstein, 2019)} has presented a geometric deep learning approach for fake news detection on Twitter social network. The proposed method allows integrating heterogeneous data pertaining to the user profile and activity, social network structure, news spreading patterns and content. The proposed model achieves  promising and robust results on the large-scale dataset, pointing to the great potential of GNN-based methods for fake news detection. Later on, GCNSI
(Graph Convolutional Networks based Source Identification) has been proposed to locate multiple rumor sources without prior knowledge of underlying propagation model. By adopting spectral domain convolution, GCNSI learns node representation by utilizing its multi-order neighbors information such that the prediction precision on the sources is improved. Moreover, Tian et.al~\hyperref[csl:138]{(Bian et al., 2020)} argue that the existing methods only take into account the patterns of deep propagation but ignore the structures of wide dispersion in disinformation (rumor) detection. They propose a novel bi-directional graph model, named Bi-Directional Graph Convolutional Networks (Bi-GCN), to explore both characteristics by operating on both top-down and bottom-up propagation of rumors. Its inherent GCN model gives the proposed method Bi-GCN the ability of processing graph/tree structures and learning higher-level representations more conducive to disinformation (rumor) detection.

\section{Mitigating Disinformation via Education, Research, and Collaboration}

In recent years, especially after the 2016 US election, disinformation has become a worry to researchers from different fields and government officials. We noticed a surge in research related to identifying, understanding, and combating disinformation. Also, we see that many governments and educational organizations around the world are trying to combat disinformation.  In this section, we discuss the research fields that are concerned about disinformation, and we list the strategies they give to combat disinformation.  In addition to that, we talk about efforts related to combating disinformation through education. 

\subsection{Mitigating Disinformation via Education}
In an alarming survey report by the Stanford History Education Group \hyperref[csl:139]{(Breakstone et al., 2019)}, over 3,400 high school students in the United States were evaluated on their ability to differentiate ``fact from fiction" and how well they were able to judge the credibility of information sources on online social media. According to the report, students were assigned six tasks that evaluated how they consume news on social media, and majority of the students failed at the tasks they were assigned, thus proving the need for administering digital media evaluation skills to students early on in their life. 

Recognizing the alarming prevalence of fake news and disinformation, and the potential threat to consumers of such false information, many countries, government and private bodies have been investing in steps to educate the general public about disinformation and how to spot fake news. Many schools across all 50 states in the US \hyperref[csl:140]{(Chen, 20AD}; \hyperref[csl:141]{Timsit, 12AD}; \hyperref[csl:142]{Tugend, 20AD)} have modified their curricula to add courses to educate students to be critical to the news they see on social media. \textit{News Literacy Project \footnote{https://newslit.org/}} - a US education non-profit - is responsible for designing the curriculum and offering nonpartisan, independent programs that teach students how to know what to trust in the digital age. Recently, Google has also launched media literacy activities to teach kids how to be aware of the information they see online, and how to identify fake posts and URLs\hyperref[csl:143]{(Mascott, 2019)}. They have also partnered with YMCA and National PTA to host workshops for children and parents.

Similar approaches to educate citizens are being taken across the globe. Finland, which ranks the highest in terms of media literacy among European nations \hyperref[csl:144]{(Lessenski, 2018)}, have schools and colleges teaching students how to evaluate the authenticity of articles on social media before they ``like" or ``share" them.

\subsection{Mitigating Disinformation via Research and Collaboration}
Many researchers from different fields are trying to understand disinformation in order to mitigate its effect on society. For instance, cognitive scientists study the problem of disinformation through studying why do people believe disinformation and what type of people are more likely to believe it \hyperref[csl:145]{(Bronstein, Pennycook, Bear, Rand, \& Cannon, 2019)}. Journalists are trying to re-evaluate the reporting of the news and how ``elites'' should discuss disinformation in the media \hyperref[csl:146]{(Van Duyn \& Collier, 2019)}. Computer scientists are applying data mining and machine learning to detect and deal with disinformation (see section \ref{sec_detection}). Here, we talk about these efforts, and we list some of the strategies that researchers are suggesting to mitigate the spread of disinformation. 

\textcolor{black}{\subsubsection{Computational Approaches}}

\color{black}
    
Given the prevalence of disinformation, websites and social media platforms have been focusing a lot of their computational resources on mitigation methods \hyperref[csl:147]{(Shu, Bernard, \& Liu, 2018)}, some of which are briefly explained below:

\textit{Source Identification}: Identifying malicious sources involved in the creation and spread of fake news and disinformation. An important step in this process is the task of information provenance - identifying the sources given the diffusion network of the news article \hyperref[csl:148]{(Barbier, Feng, Gundecha, \& Liu, 2013)}, using provenance paths. Influential social media users having a large online follower-base may also be responsible for the rapid spread of disinformation. Identifying these ``opinion leaders'' propagating disinformation and terminating their accounts or even slowing down their reach or may slow down the spread of disinformation. 

\textit{Network Intervention}: Various algorithmic techniques have been developed to slow down and limit the spread of disinformation. Most of the techniques are based on the concept of diffusion of information on social network of users, following the dynamics of Independent Cascade Model or Linear Threshold Model. Researchers tackling this issue have proposed techniques like Influence Minimization or Influence Limitation. Some of these approaches focus on designing cascades of counter campaigns such that the disinformation cascade slows down \hyperref[csl:149]{(Budak, Agrawal, \& Abbadi, 2011)} \hyperref[csl:150]{(Tong, Du, \& Wu, 2018)}, while some others focus on finding the smallest set of nodes to remove, such that the influence is minimized \hyperref[csl:151]{(Pham, Phu, Hoang, Pei, \& Thai, 2019)}.

\textit{Content Flagging}: Social media platforms often provide users the option to `flag' a post as spreading false information, and once a post gets sufficient number of these `flags', it is often fact-checked, and if found to be disinformation, it is subsequently removed from the platform. Given the trade-off between the number of user flags and the potential damage caused by increased exposure to disinformation, Kim et al. \hyperref[csl:152]{(Kim, Tabibian, Oh, Sch{\"o}lkopf, \& Gomez-Rodriguez, 2018)} proposed a scalable algorithm to effectively select articles for fact-checking and and schedule the fact-checking process.

\color{black}

\subsubsection{Cognitive Science}

Cognitive science is defined as the scientific study of the human mind to understand how human knowledge used, processed, and acquired \footnote{https://bcs.mit.edu/research/cognitive-science}. Cognitive scientists are concerned about disinformation, and they are applying their knowledge to find how and why people believe disinformation. Bronstein et al. \hyperref[csl:145]{(Bronstein, Pennycook, Bear, Rand, \& Cannon, 2019)} studied the reasons that might lead people to believe in fake news. They found that the main factors behind believing disinformation are Delusionality, Dogmatism, Religious Fundamentalism, and Reduced Analytic Thinking. Bronstein et al. concluded that people who uphold delusion-like beliefs, dogmatic individuals, or religious fundamentalists are more likely to believe disinformation. They argue that the reason behind this phenomenon is that these people are less likely to engage in open-minded thinking and analytic thinking, which makes them vulnerable to disinformation. Pennycook et al. \hyperref[csl:153]{(Pennycook \& Rand, 2019)} found that believing disinformation is more related to lazy thinking or lack of it more than partisan bias. 

The lack of thinking and other factors that lead to believing disinformation is a solvable problem.  De keersmaecker and Roets \hyperref[csl:154]{(Keersmaecker \& Roets, 2017)} demonstrate that, generally, people adjust their attitudes towards disinformation once they are faced with the truth. However, the level of cognitive ability influences the degree to which people change their assessment of the news. Cognitive scientist believes that leveraging interventions that improve analytic and actively open-minded thinking might lessen the belief in disinformation \hyperref[csl:145]{(Bronstein, Pennycook, Bear, Rand, \& Cannon, 2019}; \hyperref[csl:153]{Pennycook \& Rand, 2019)}. Actively open-minded thinking ``involves the search for alternative explanations and the use of evidence to revise beliefs.'' Furthermore, analytic thinking is ``the disposition to initiate deliberate thought processes in order to reflect on intuitions and gut feelings.'' \hyperref[csl:145]{(Bronstein, Pennycook, Bear, Rand, \& Cannon, 2019)}

\subsubsection{Journalism and Political Science}

Following the 2016 US election, many political scientists and journalists studied the effect of disinformation on social media on society. Also, they studied disinformation to find the people that are more likely to be victims of fake news to develop strategies to mitigate the problem of disinformation. Bovet et al. \hyperref[csl:155]{(Bovet \& Makse, 2019)} found that confirmation bias and social influence is the main reason for echo chambers where users with similar beliefs share disinformation about a specific topic.  However, users who are exposed to disinformation is a small fraction of the general public \hyperref[csl:156]{(Grinberg, Joseph, Friedland, Swire-Thompson, \& Lazer, 2019)}.  Those users tend to be conservative-leaning, older, and highly engaged with political news \hyperref[csl:156]{(Grinberg, Joseph, Friedland, Swire-Thompson, \& Lazer, 2019)}.  Those users tend to consume disinformation about political news  \hyperref[csl:156]{(Grinberg, Joseph, Friedland, Swire-Thompson, \& Lazer, 2019)} or science \hyperref[csl:55]{(Scheufele \& Krause, 2019)}. 

Journalists and political scientists wanted to study the reasons behind the distrust of traditional media. Van Duyn et al. \hyperref[csl:146]{(Van Duyn \& Collier, 2019)} wanted to study the effect of elite discourse on disinformation and how does it affect people's trust of news organization.  They found that people exposed to the elite discussion about disinformation tend to have lower levels of trust in traditional media, and they have a hard time identifying real information.  Also, attacking real news organizations, and branding them as fake news sites lead to the rise of disinformation on social media \hyperref[csl:157]{(Tandoc Jr, 2019)}.

Journalists and political scientists suggest that users should consume news from verified accounts such as well known journalists or official accounts such as government accounts \hyperref[csl:155]{(Bovet \& Makse, 2019)}.  In addition to that, disinformation experts should be careful about the way they discuss disinformation in order to avoid causing distrust in the media in general \hyperref[csl:146]{(Van Duyn \& Collier, 2019)}. Politicians should also stop attacking the real news media and label them as fake news.

\section{Conclusion and Future Work}
In this paper, we gave an overview of disinformation from different aspects. We defined disinformation, and we listed some of the forms it takes, such as fake news, rumors, hoaxes, etc. We also talked about the history of disinformation and how internet and social media affects the spread and subsequent consumption of disinformation. We discuss the challenges in detecting fake news and then proceed to talk about the different forms of fabricated content that may exist on social media, i.e. text, image, video etc., and discuss the notable detection techniques devised by researchers for detecting the different forms of fabricated content. We also talked about the ongoing efforts to inform people about presence of disinformation, in terms of educational programs, and also gave a brief perspective into disinformation through the lenses of cognitive science and political science.
The main goal of our work is to provide the reader with a comprehensive view of the current scenario in the research field of disinformation detection. Given the ever-growing relevance of disinformation on social media, and the complementary need for better detection methods and models, we hope our work would provide future researchers an idea of the advances made so far, and hence guide them towards further improvement. In this regard, we also list some of the datasets and tools used most widely in the context of disinformation detection. 

There are several interesting and promising directions to explore. First, threat modeling of disinformation is an exciting topic to study. Threat modeling is a widely used technique in the field of computer security to identify and combat the threat. We believe that threat modeling is a very interesting topic that was not studied to its full potential. Disinformation generators could be developed as threat models to improve our understanding of disinformation and how it spreads in society. Building threat models may help to build better disinformation detectors. To the best of our knowledge, there is no work on disinformation threat modeling other than Zellers et al. \hyperref[csl:119]{(Zellers et al., 2019)}. However, Zellers et al. \hyperref[csl:119]{(Zellers et al., 2019)} work is limited to only textual data. As we explained previously, real-world disinformation comes in many forms, such as text, images, and videos. To build better threats models, these models need to be able to generate disinformation in as many forms as possible. 

%One of the least researched forms of disinformation is video.

Now that generating deepfakes and fake videos in general has become extremely easy, we feel there is a need for more work to be done in the detection of such fabricated videos. There is a great need for video threat modeling. One possible way to create these models is by using stock video databases available online. This method is inexpensive and quick to overwhelm detectors. %Also, creating a massive number of disinformation videos increase the probability that one of them becomes viral and cause more damage. 
Moreover, very few labeled deepfake datasets are available publicly for research. Work could be done in generating and creating more annotated deepfake datasets.

%Many disinformation detectors rely on users' responses. Most of them rely on user sentiment or stance, as detected via user comments. A good threat model might generate fake responses to trick these detectors, as well as real users. 

The way that social media users react to novel news content is an excellent way to detect disinformation. People tend to respond differently to disinformation and real information. However, some people are more likely to get fooled by disinformation than others. The reason behind this phenomena might be confirmation bias or some other cognitive or psychological reasons, and more research could be done to study disinformation from a psychological point of view. People who get fooled by disinformation easily could be used as an indicator of disinformation. Also, those people tend to form echo chambers, and they share disinformation among themselves. Therefore, we believe that tackling the problem of echo chambers could possibly help early detection. 

\color{black}

As we mentioned previously, the problem of early detection of fake news on social media platforms is an unsolved one and much improvements could be done in this regard. Furthermore, the problem of fake news is extremely topic-dependent. Hence, in case of ML-based methods, models trained using data belong to a certain domain or set of events may fail to perform satisfactorily in a different context. To tackle this challenge, along with the challenge of scarcity of available data when it comes to novel fake news, transfer learning based techniques could be used. Newly available data could be used to further tune a model that has already learned the linguistic structures of fake vs. real news. 

\color{black}

\appendix
\section{Datasets and Tools Available}

%% by amrita

With the gradually increasing volume of ongoing research in the domain of fake news or disinformation detection, there are a considerable number of datasets for use by researchers.  

\subsection{Datasets}

There have been a few small fake news datasets predating the era during which research in fake news detection gained momentum. The use of advanced computational methods and machine learning frameworks engenders the need for large datasets that can be used to train such automatic detection models in order to achieve improved performance. Here we describe a few datasets that have been in use recently:

\textbf{[Twitter Media Corpus \footnote{\url{https://github.com/MKLab-ITI/image-verification-corpus}}]}: This dataset consists of tweets from Twitter containing multimedia content and associated `fake' or `real' labels. Tweets where such fake images were used as supporting media have been labelled as fake. For more information on the dataset and the framework developed by the creators of the dataset, please refer to \hyperref[csl:158]{(Boididou et al., 2018)}.

\textbf{[PHEME Rumor Dataset \footnote{\url{https://figshare.com/articles/PHEME\_dataset\_for\_Rumour\_Detection\_and\_Veracity\_Classification/6392078}}]}: This dataset \hyperref[csl:159]{(Kochkina, Liakata, \& Zubiaga, 2018)} contains a collection of tweets from Twitter, comprising of both rumors and non-rumors. These tweets correspond to 9 different events, and are annotated with labels - True, False or Unverified. For details of the data collection and the related analysis by the authors, please refer to \hyperref[csl:160]{(Zubiaga, Liakata, Procter, Hoi, \& Tolmie, 2016)}.

\textbf{[CREDBANK \footnote{\url{https://github.com/compsocial/CREDBANK-data}}]}: This is a dataset \hyperref[csl:161]{(Mitra \& Gilbert, 2015)} containing over 60 million tweets from Twitter, over 1049 real-world events. As explained in the paper introducing this dataset, each of the tweets have been annotated with a credibility score as judged by human annotator. The credibility score is based on a credibilty scale: \textit{Certainly Accurate}, \textit{Probably Accurate}, \textit{Uncertain}, \textit{Probably Inaccurate}, and \textit{Certainly Inaccurate}.

\textbf{[LIAR \footnote{\url{https://www.cs.ucsb.edu/\~william/data/liar\_dataset.zip}}]}: Consists of 12.8K labelled sentences across different contexts, extracted from {\tt POLITIFACT.COM}. The sentences consist of statements made by speakers having different political affiliations and also Facebook posts. Each statement is associated with a fine-grained truthfulness label, which is one of 6 possible values. Furthermore, each speaker is associated with a `credit history', that is a tuple having counts of how many statements they made in each of the truthfulness category, thus serving as some sort of credibility score. More information regarding the dataset can be found in \hyperref[csl:162]{(Wang, 2017)}.

\textbf{[Twitter Death Hoax \footnote{\url{https://figshare.com/articles/Twitter\_Death\_Hoaxes\_dataset/5688811}}]}: This dataset consists of death reports posted on Twitter between 1st January, 2012 and 31st December 2014. There are over 4k reports, out of which only 2031 are real deaths, as verified by using information form Wikipedia. For more details on the collection and and related hoax detection task, please refer to \hyperref[csl:163]{(Zubiaga, 2018)}.

\textbf{[FA-KES: A Fake News Dataset around the Syrian War \footnote{\url{ https://doi.org/10.5281/zenodo.2607278}}]}: This dataset consists of news articles focused on the Syrian war and these articles are labelled as ``fake'' or ``credible'' by comparing the content of the articles with ground truth obtained from Syrian Violations Documentation Center. For more information regarding the data collection and annotation process, please refer to \hyperref[csl:164]{(Salem, Al Feel, Elbassuoni, Jaber, \& Farah, 2019)}.

% \textbf{[Weibo]}: Sina Weibo is a Chinese microblogging website, which has often been referred to as the Chinese equivalent of Twitter. Although this dataset is not publicly available, to the best of our knowledge, there have been several works using fake vs real news data scraped from Weibo, offering insight on ways to scrape the website and gather data. For details on how to scrape Weibo data and create your own dataset, please refer to \hyperref[csl:165]{(Littman, Kerchner, He, Tan, \& Zeljak, 2017)}.

\textbf{[FakeNewsNet \footnote{\url{https://github.com/KaiDMML/FakeNewsNet}}]} : Consists of real and fake news articles related to politics (from {\tt POLITIFACT} and celebrities (from {\tt GOSSIPCOP}. Along with the content of the posts/tweets, the dataset also has data related to the social context - including user-profiles, user-following, user-followers etc. For more details, please refer to \hyperref[csl:166]{(Shu, Mahudeswaran, Wang, Lee, \& Liu, 2018)}.

\subsection{Tools}
%by Amrita
Given the significant  amount of ongoing work in the field of fake news or disinformation detection, researchers and practitioners have made certain tools available for use, in order to further the research in this area. Here we list a few publicly available ones:

\textbf{[Big Bird \footnote{\url{https://bigbird.dev/}}]}: This is a tool to automatically generate (fake) news articles. The website {\tt www.notrealnews.net} contain `news' generated by this text generator tool and subsequently edited by human agents. 

\textbf{[Computational-Verification \footnote{\url{https://github.com/MKLab-ITI/computational-verification}}]}: This is a framework for identifying the veracity of content/posts on social media (Twitter, in this case). This makes sue of two types of features - \textit{Tweet-based features}, such as number of words, number of retweets etc, and \textit{User-based features} such as friend-follower ratio, number of tweets etc. A two-level classification model, to identify fake vs. real content, is trained. For more details and information on the approach, please refer \hyperref[csl:158]{(Boididou et al., 2018)}.

\textbf{[Facebook Hoax \footnote{\url{https://github.com/gabll/some-like-it-hoax/tree/master/dataset}}]}: This is a collection of Python scripts to collect posts on a page on Facebook, since a pre-defined date as input by the user. Authors in \hyperref[csl:167]{(Tacchini, Ballarin, Della Vedova, Moret, \& de Alfaro, 2017)} use this to create a dataset of over 15,000 posts on Facebook and try to identify hoax vs. non-hoax posts based on the users who `liked' the posts. The authors used logistic regression and harmonic boolean label crowdsourcing in order to perform the classification. For more details please refer to \hyperref[csl:167]{(Tacchini, Ballarin, Della Vedova, Moret, \& de Alfaro, 2017)}. 

% \textbf{[Hoaxy \footnote{\url{https://hoaxy.iuni.iu.edu/}}]}: This is a tool to collect and track disinformation in the form of tweets.

\selectlanguage{english}
\FloatBarrier
\section*{References}\sloppy
\phantomsection
\label{csl:63} (n.d.). . \url{https://thetrustproject.org/.} Retrieved from \url{https://thetrustproject.org/}

\phantomsection
\label{csl:64} (n.d.). . \url{https://infogram.com/politifacts-fake-news-almanac-1gew2vjdxl912nj.}

\phantomsection
\label{csl:65} (n.d.). . \url{https://mediabiasfactcheck.com/.} Retrieved from \url{https://mediabiasfactcheck.com/}

\phantomsection
\label{csl:62}Abbasi, M.-A., \& Liu, H. (2013). {Measuring user credibility in social media}. In \textit{International Conference on Social Computing, Behavioral-Cultural Modeling, and Prediction} (pp. 441–448). Springer.

\phantomsection
\label{csl:88}Abokhodair, N., Yoo, D., \& McDonald, D. W. (2015). {Dissecting a social botnet: Growth, content and influence in Twitter}. In \textit{Proceedings of the 18th ACM Conference on Computer Supported Cooperative Work {\&} Social Computing} (pp. 839–851). ACM.

\phantomsection
\label{csl:6}Allcott, H., \& Gentzkow, M. (2017). \textit{{Social Media and Fake News in the 2016 Election}}. National Bureau of Economic Research. Retrieved from \url{https://doi.org/10.3386\%2Fw23089}

\phantomsection
\label{csl:97}Allem, J.-P., Ferrara, E., Uppu, S. P., Cruz, T. B., \& Unger, J. B. (2017). {E-Cigarette Surveillance With Social Media Data: Social Bots, Emerging Topics, and Trends}. \textit{JMIR Public Health and Surveillance}, \textit{3}(4), e98. Retrieved from \url{http://publichealth.jmir.org/2017/4/e98/}

\phantomsection
\label{csl:71}Allport, G. W., \& Postman, L. (1947). {The psychology of rumor.}. Oxford, England: Henry Holt.

\phantomsection
\label{csl:76}Anthony, S. (1973). {Anxiety and rumor}. \textit{The Journal of social psychology}, \textit{89}(1), 91–98. Taylor \& Francis.

\phantomsection
\label{csl:23}Arun, C. (2019). {On WhatsApp, Rumours, and Lynchings}. \textit{Economic \& Political Weekly}, \textit{54}(6), 30–35.

\phantomsection
\label{csl:148}Barbier, G., Feng, Z., Gundecha, P., \& Liu, H. (2013). {Provenance Data in Social Media}. \textit{Synthesis Lectures on Data Mining and Knowledge Discovery}, \textit{4}(1), 1–84. Morgan {\&} Claypool Publishers {LLC}. Retrieved from \url{https://doi.org/10.2200\%2Fs00496ed1v01y201304dmk007}

\phantomsection
\label{csl:92}Bessi, A., \& Ferrara, E. (2016). {Social bots distort the 2016 U.S. Presidential election online discussion}. \textit{First Monday}, \textit{21}(11). Retrieved from \url{https://uncommonculture.org/ojs/index.php/fm/article/view/7090}

\phantomsection
\label{csl:32}Bettencourt, L. M. A., Cintr{\'o}n-Arias, A., Kaiser, D. I., \& Castillo-Ch{\'a}vez, C. (2006). {The power of a good idea: Quantitative modeling of the spread of ideas from epidemiological models}. \textit{Physica A: Statistical Mechanics and its Applications}, \textit{364}, 513–536. Elsevier.

\phantomsection
\label{csl:138}Bian, T., Xiao, X., Xu, T., Zhao, P., Huang, W., Rong, Y., \& Huang, J. (2020). {Rumor Detection on Social Media with Bi-Directional Graph Convolutional Networks}. \textit{arXiv preprint arXiv:2001.06362}.

\phantomsection
\label{csl:19}Boghardt, T. (2009). {Soviet Bloc intelligence and its AIDS disinformation campaign}. \textit{Studies in Intelligence}, \textit{53}(4), 1–24.

\phantomsection
\label{csl:158}Boididou, C., Papadopoulos, S., Zampoglou, M., Apostolidis, L., Papadopoulou, O., \& Kompatsiaris, Y. (2018). {Detection and visualization of misleading content on Twitter}. \textit{International Journal of Multimedia Information Retrieval}, \textit{7}(1), 71–86. Springer.

\phantomsection
\label{csl:155}Bovet, A., \& Makse, H. A. (2019). {Influence of fake news in Twitter during the 2016 US presidential election}. \textit{Nature communications}, \textit{10}(1), 1–14. Nature Publishing Group.

\phantomsection
\label{csl:139}Breakstone, J., Smith, M., Wineburg, S., Rapaport, A., Carle, J., Garland, M., \& Saavedra, A. (2019). {Students’ civic online reasoning: A national portrait}. Stanford History Education Group \& Gibson Consulting.

\phantomsection
\label{csl:96}Broniatowski, D. A., Jamison, A. M., Qi, S. H., AlKulaib, L., Chen, T., Benton, A., Quinn, S. C., et al. (2018). {Weaponized Health Communication: Twitter Bots and Russian Trolls Amplify the Vaccine Debate}. \textit{American Journal of Public Health}, \textit{108}(10), 1378–1384. Retrieved from \url{https://ajph.aphapublications.org/doi/pdf/10.2105/AJPH.2018.304567}

\phantomsection
\label{csl:145}Bronstein, M. V., Pennycook, G., Bear, A., Rand, D. G., \& Cannon, T. D. (2019). {Belief in fake news is associated with delusionality, dogmatism, religious fundamentalism, and reduced analytic thinking}. \textit{Journal of Applied Research in Memory and Cognition}, \textit{8}(1), 108–117. Elsevier.

\phantomsection
\label{csl:149}Budak, C., Agrawal, D., \& Abbadi, A. E. (2011). {Limiting the spread of misinformation in social networks}. In \textit{Proceedings of the 20th international conference on World wide web - {WWW} {\textquotesingle}11}. {ACM} Press. Retrieved from \url{https://doi.org/10.1145\%2F1963405.1963499}

\phantomsection
\label{csl:101}Cao, Q., Sirivianos, M., Yang, X., \& Pregueiro, T. (2012). {Aiding the detection of fake accounts in large scale social online services}. In \textit{Proceedings of the 9th USENIX conference on Networked Systems Design and Implementation} (p. 15). USENIX Association.

\phantomsection
\label{csl:115}Castillo, C., El-Haddad, M., Pfeffer, J., \& Stempeck, M. (2014). {Characterizing the life cycle of online news stories using social media reactions}. In \textit{Proceedings of the 17th ACM conference on Computer supported cooperative work \& social computing} (pp. 211–223).

\phantomsection
\label{csl:140}Chen, S. (20AD). {Schools around the world are now teaching kids to spot fake news}. \textit{Quartz}. Retrieved from \url{https://qz.com/1175155/a-special-class-how-to-teach-kids-to-spot-fake-news/}

\phantomsection
\label{csl:44}Chollet, F. (2017). {Xception: Deep learning with depthwise separable convolutions}. In \textit{Proceedings of the IEEE conference on computer vision and pattern recognition} (pp. 1251–1258).

\phantomsection
\label{csl:108}Chu, Z., Gianvecchio, S., Wang, H., \& Jajodia, S. (2012). {Detecting automation of Twitter accounts: Are you a human, bot, or cyborg?}. \textit{IEEE Transactions on Dependable and Secure Computing}, \textit{9}(6), 811–824. IEEE.

\phantomsection
\label{csl:125}Ciampaglia, G. L., Shiralkar, P., Rocha, L. M., Bollen, J., Menczer, F., \& Flammini, A. (2015). {Computational fact checking from knowledge networks}. \textit{PloS one}, \textit{10}(6), e0128193. Public Library of Science San Francisco, CA USA.

\phantomsection
\label{csl:54}Cui, L., \& Lee, S. W. D. (2019). {SAME: Sentiment-Aware Multi-Modal Embedding for Detecting Fake News}.

\phantomsection
\label{csl:20}De Maeyer, D. (1997). {Internet’s information highway potential}. \textit{Internet Research}. MCB UP Ltd.

\phantomsection
\label{csl:68}DiFonzo, N., \& Bordia, P. (2007). \textit{{Rumor psychology: Social and organizational approaches.}}. American Psychological Association.

\phantomsection
\label{csl:135}Ding, K., Li, J., Bhanushali, R., \& Liu, H. (2019). {Deep anomaly detection on attributed networks}. In \textit{Proceedings of the 2019 SIAM International Conference on Data Mining} (pp. 594–602). SIAM.

\phantomsection
\label{csl:136}Ding, K., Li, Y., Li, J., Liu, C., \& Liu, H. (2019). {Graph Neural Networks with High-order Feature Interactions}. \textit{arXiv preprint arXiv:1908.07110}.

\phantomsection
\label{csl:79}Elder, J. (2013, November). {Inside a Twitter Robot Factory; Fake Activity, Often Bought for Publicity Purposes, Influences Trending Topics}. \textit{Wall Street Journal (Online)}. Retrieved from \url{https://www.wsj.com/articles/bogus-accounts-dog-twitter-1385335134}

\phantomsection
\label{csl:123}Esmaeilzadeh, S., Peh, G. X., \& Xu, A. (2019). {Neural Abstractive Text Summarization and Fake News Detection}. \textit{arXiv preprint arXiv:1904.00788}.

\phantomsection
\label{csl:74}Fernandez, M., Alvarez, L., \& Nixon, R. (2017, October 22). {Still Waiting for FEMA in Texas and Florida After Hurricanes}. The New York Times. Retrieved from \url{https://www.nytimes.com/2017/10/22/us/fema-texas-florida-delays-.html}

\phantomsection
\label{csl:100}Ferrara, E., Varol, O., Davis, C., Menczer, F., \& Flammini, A. (2016). {The rise of social bots}. \textit{Communications of the ACM}, \textit{59}(7), 96–104. ACM.

\phantomsection
\label{csl:4}Fetzer, J. H. (2004). {Disinformation: The use of false information}. \textit{Minds and Machines}, \textit{14}(2), 231–240. Springer.

\phantomsection
\label{csl:49}George, S. (13AD). {`Deepfakes' called new election threat, with no easy fix}. \textit{AP News}.

\phantomsection
\label{csl:33}Goldenberg, J., Libai, B., \& Muller, E. (2001). {Using complex systems analysis to advance marketing theory development: Modeling heterogeneity effects on new product growth through stochastic cellular automata}. \textit{Academy of Marketing Science Review}, \textit{9}(3), 1–18.

\phantomsection
\label{csl:34}Goldenberg, J., Libai, B., \& Muller, E. (2001). {Talk of the network: A complex systems look at the underlying process of word-of-mouth}. \textit{Marketing letters}, \textit{12}(3), 211–223. Springer.

\phantomsection
\label{csl:39}Goodfellow, I., Pouget-Abadie, J., Mirza, M., Xu, B., Warde-Farley, D., Ozair, S., Courville, A., et al. (2014). {Generative adversarial nets}. In \textit{Advances in neural information processing systems} (pp. 2672–2680).

\phantomsection
\label{csl:35}Granovetter, M. (1978). {Threshold models of collective behavior}. \textit{American journal of sociology}, \textit{83}(6), 1420–1443. University of Chicago Press.

\phantomsection
\label{csl:73}Grenoble, R. (2017, July 9). {Hurricane Harvey Is Just The Latest In Facebook’s Fake News Problem}. Huffington Post. Retrieved from \url{https://www.huffingtonpost.com/entry/facebook-hurricane-harvey-fake-news_us_59b17900e4b0354e441021fb}

\phantomsection
\label{csl:156}Grinberg, N., Joseph, K., Friedland, L., Swire-Thompson, B., \& Lazer, D. (2019). {Fake news on Twitter during the 2016 US presidential election}. \textit{Science}, \textit{363}(6425), 374–378. American Association for the Advancement of Science.

\phantomsection
\label{csl:111}Guacho, G. B., Abdali, S., Shah, N., \& Papalexakis, E. E. (2018). {Semi-supervised Content-Based Detection of Misinformation via Tensor Embeddings}. In \textit{2018 {IEEE}/{ACM} International Conference on Advances in Social Networks Analysis and Mining ({ASONAM})}. {IEEE}. Retrieved from \url{https://doi.org/10.1109\%2Fasonam.2018.8508241}

\phantomsection
\label{csl:66}Guess, A., Nyhan, B., \& Reifler, J. (2018). {Selective Exposure to Misinformation: Evidence from the consumption of fake news during the 2016 US presidential campaign}. Retrieved from \url{https://www.dartmouth.edu/~nyhan/fake-news-2016.pdf}

\phantomsection
\label{csl:50}G{\"u}era, D., \& Delp, E. J. (2018). {Deepfake video detection using recurrent neural networks}. In \textit{2018 15th IEEE International Conference on Advanced Video and Signal Based Surveillance (AVSS)} (pp. 1–6). IEEE.

\phantomsection
\label{csl:58}Haller, A., \& Holt, K. (2019). {Paradoxical populism: How PEGIDA relates to mainstream and alternative media}. \textit{Information, Communication \& Society}, \textit{22}(12), 1665–1680. Taylor \& Francis.

\phantomsection
\label{csl:56}Hasher, L., Goldstein, D., \& Toppino, T. (1977). {Frequency and the conference of referential validity}. \textit{Journal of verbal learning and verbal behavior}, \textit{16}(1), 107–112. Elsevier.

\phantomsection
\label{csl:12}Hernandez, J. C., Hernandez, C. J., Sierra, J. M., \& Ribagorda, A. (2002). {A first step towards automatic hoax detection}. In \textit{Proceedings. 36th Annual 2002 International Carnahan Conference on Security Technology}. {IEEE}. Retrieved from \url{https://doi.org/10.1109\%2Fccst.2002.1049234}

\phantomsection
\label{csl:5}Hernon, P. (1995). {Disinformation and misinformation through the internet: Findings of an exploratory study}. \textit{Government Information Quarterly}, \textit{12}(2), 133–139. Elsevier {BV}. Retrieved from \url{https://doi.org/10.1016\%2F0740-624x\%2895\%2990052-7}

\phantomsection
\label{csl:59}Hirning, N. P., Chen, A., \& Shankar, S. (2017). \textit{{Detecting and identifying bias-heavy sentences in news articles}}. Technical report, Stanford University.

\phantomsection
\label{csl:112}Hosseinimotlagh, S., \& Papalexakis, E. E. (2018). {Unsupervised content-based identification of fake news articles with tensor decomposition ensembles}. \textit{Proceedings of the Workshop on Misinformation and Misbehavior Mining on the Web (MIS2)}.

\phantomsection
\label{csl:60}Iyyer, M., Enns, P., Boyd-Graber, J., \& Resnik, P. (2014). {Political ideology detection using recursive neural networks}. In \textit{Proceedings of the 52nd Annual Meeting of the Association for Computational Linguistics (Volume 1: Long Papers)} (pp. 1113–1122).

\phantomsection
\label{csl:75}Janosch Delcker, Z. W., \& Scott, M. (2020). {The coronavirus fake news pandemic sweeping WhatsApp}. \textit{Politico}. \url{https://www.politico.com/news/2020/03/16/coronavirus-fake-news-pandemic-133447.} Retrieved from \url{https://www.politico.com/news/2020/03/16/coronavirus-fake-news-pandemic-133447}

\phantomsection
\label{csl:31}Jin, F., Dougherty, E., Saraf, P., Cao, Y., \& Ramakrishnan, N. (2013). {Epidemiological modeling of news and rumors on twitter}. In \textit{Proceedings of the 7th Workshop on Social Network Mining and Analysis} (p. 8). ACM.

\phantomsection
\label{csl:133}Jin, Z., Cao, J., Guo, H., Zhang, Y., Wang, Y., \& Luo, J. (2017). {Detection and analysis of 2016 us presidential election related rumors on twitter}. In \textit{International conference on social computing, behavioral-cultural modeling and prediction and behavior representation in modeling and simulation} (pp. 14–24). Springer.

\phantomsection
\label{csl:128}Jin, Z., Cao, J., Zhang, Y., \& Luo, J. (2016). {News verification by exploiting conflicting social viewpoints in microblogs}. In \textit{Thirtieth AAAI conference on artificial intelligence}.

\phantomsection
\label{csl:120}Jozefowicz, R., Vinyals, O., Schuster, M., Shazeer, N., \& Wu, Y. (2016). {Exploring the limits of language modeling}. \textit{arXiv preprint arXiv:1602.02410}.

\phantomsection
\label{csl:154}Keersmaecker, J. D., \& Roets, A. (2017). {`Fake news': Incorrect but hard to correct. The role of cognitive ability on the impact of false information on social impressions}. \textit{Intelligence}, \textit{65}, 107–110. Elsevier {BV}. Retrieved from \url{https://doi.org/10.1016\%2Fj.intell.2017.10.005}

\phantomsection
\label{csl:85}Kelion, L., \& Silva, S. (2016). {Pro-Clinton bots 'fought back but outnumbered in second debate'}. BBC News. Retrieved from \url{http://www.bbc.com/news/technology-37703565}

\phantomsection
\label{csl:87}Khaund, T., Al-Khateeb, S., Tokdemir, S., \& Agarwal, N. (2018). {Analyzing Social Bots and Their Coordination During Natural Disasters}. In \textit{International Conference on Social Computing, Behavioral-Cultural Modeling and Prediction and Behavior Representation in Modeling and Simulation} (pp. 207–212). Springer.

\phantomsection
\label{csl:30}Kim, J., Tabibian, B., Oh, A., Sch{\"o}lkopf, B., \& Gomez-Rodriguez, M. (2018). {Leveraging the crowd to detect and reduce the spread of fake news and misinformation}. In \textit{Proceedings of the Eleventh ACM International Conference on Web Search and Data Mining} (pp. 324–332). ACM.

\phantomsection
\label{csl:152}Kim, J., Tabibian, B., Oh, A., Sch{\"o}lkopf, B., \& Gomez-Rodriguez, M. (2018). {Leveraging the crowd to detect and reduce the spread of fake news and misinformation}. In \textit{Proceedings of the Eleventh ACM International Conference on Web Search and Data Mining} (pp. 324–332).

\phantomsection
\label{csl:134}Kipf, T. N., \& Welling, M. (2016). {Semi-supervised classification with graph convolutional networks}. \textit{arXiv preprint arXiv:1609.02907}.

\phantomsection
\label{csl:91}Kitzie, V. L., Karami, A., \& Mohammadi, E. (2018). {``Life Never Matters in the Democrats Mind``: Examining Strategies of Retweeted Social Bots During a Mass Shooting Event}. \textit{arXiv:1808.09325}. Retrieved from \url{http://arxiv.org/pdf/1808.09325v1}

\phantomsection
\label{csl:159}Kochkina, E., Liakata, M., \& Zubiaga, A. (2018). {PHEME dataset for Rumour Detection and Veracity Classification}. Retrieved from \url{https://figshare.com/articles/PHEME_dataset_for_Rumour_Detection_and_Veracity_Classification/6392078}

\phantomsection
\label{csl:80}Koh, Y. (2014, March). {Only 11{\%} of New Twitter Users in 2012 Are Still Tweeting}. \textit{Dow Jones Institutional News}. Retrieved from \url{https://blogs.wsj.com/digits/2014/03/21/new-report-spotlights-twitters-retention-problem/}

\phantomsection
\label{csl:107}Kudugunta, S., \& Ferrara, E. (2018). {Deep neural networks for bot detection}. \textit{Information Sciences}, \textit{467}, 312–322.

\phantomsection
\label{csl:10}Kumar, S., West, R., \& Leskovec, J. (2016). {Disinformation on the Web}. In \textit{Proceedings of the 25th International Conference on World Wide Web - {WWW} {\textquotesingle}16}. {ACM} Press. Retrieved from \url{https://doi.org/10.1145\%2F2872427.2883085}

\phantomsection
\label{csl:124}Lakshmanan, L. V. S., Simpson, M., \& Thirumuruganathan, S. (2019). {Combating fake news: a data management and mining perspective}. \textit{Proceedings of the VLDB Endowment}, \textit{12}(12), 1990–1993. VLDB Endowment.

\phantomsection
\label{csl:77}Lazer, D. M. J., Baum, M. A., Benkler, Y., Berinsky, A. J., Greenhill, K. M., Menczer, F., Metzger, M. J., et al. (2018). {The science of fake news}. \textit{Science}, \textit{359}(6380), 1094–1096. Retrieved from \url{http://www.sciencemag.org/lookup/doi/10.1126/science.aao2998}

\phantomsection
\label{csl:48}Lee, D. (3AD). {Deepfakes porn has serious consequences}. \textit{BBC News}.

\phantomsection
\label{csl:81}Lee, K., Eoff, B. D., \& Caverlee, J. (2011). {Seven Months with the Devils: A Long-Term Study of Content Polluters on Twitter}. In \textit{ICWSM} (pp. 185–192). AAAI. Retrieved from \url{https://www.aaai.org/ocs/index.php/ICWSM/ICWSM11/paper/view/2780}

\phantomsection
\label{csl:105}Lee, S., \& Kim, J. (2014). {Early filtering of ephemeral malicious accounts on Twitter}. \textit{Computer Communications}, \textit{54}, 48–57. Elsevier.

\phantomsection
\label{csl:144}Lessenski, M. (2018). {Common Sense Wanted: Resilience to ‘Post-Truth’and its Predictors in the New Media Literacy Index 2018}. \textit{Open Society Institute, mar}.

\phantomsection
\label{csl:41}Li, Y., Chang, M.-C., \& Lyu, S. (2018). {In ictu oculi: Exposing ai generated fake face videos by detecting eye blinking}. \textit{arXiv preprint arXiv:1806.02877}.

\phantomsection
\label{csl:165}Littman, J., Kerchner, D., He, Y., Tan, Y., \& Zeljak, C. (2017). {Collecting Social Media Data from the Sina Weibo Api}. \textit{Journal of East Asian Libraries}, \textit{2017}(165), 12.

\phantomsection
\label{csl:131}Liu, Y., \& Wu, Y.-F. B. (2018). {Early detection of fake news on social media through propagation path classification with recurrent and convolutional networks}. In \textit{Thirty-Second AAAI Conference on Artificial Intelligence}.

\phantomsection
\label{csl:17}Manning, M. J., Manning, M., \& Romerstein, H. (2004). \textit{{Historical Dictionary of American Propaganda}}. Greenwood Publishing Group.

\phantomsection
\label{csl:70}Marcus, G. (2017). {How Affective Intelligence can help us understand politics}. \textit{Emotion Researcher}.

\phantomsection
\label{csl:42}Marra, F., Gragnaniello, D., Cozzolino, D., \& Verdoliva, L. (2018). {Detection of GAN-generated fake images over social networks}. In \textit{2018 IEEE Conference on Multimedia Information Processing and Retrieval (MIPR)} (pp. 384–389). IEEE.

\phantomsection
\label{csl:143}Mascott, A. (2019). {Helping kids learn to evaluate what they see online}. \url{https://www.blog.google/technology/families/be-internet-awesome-media-literacy/.} Retrieved from \url{https://blog.google/technology/families/be-internet-awesome-media-literacy/}

\phantomsection
\label{csl:161}Mitra, T., \& Gilbert, E. (2015). {Credbank: A large-scale social media corpus with associated credibility annotations}. In \textit{Ninth International AAAI Conference on Web and Social Media}. Retrieved from \url{https://github.com/compsocial/CREDBANK-data}

\phantomsection
\label{csl:137}Monti, F., Frasca, F., Eynard, D., Mannion, D., \& Bronstein, M. M. (2019). {Fake News Detection on Social Media using Geometric Deep Learning}. \textit{arXiv preprint arXiv:1902.06673}.

\phantomsection
\label{csl:106}Morstatter, F., Wu, L., Nazer, T. H., Carley, K. M., \& Liu, H. (2016). {A new approach to bot detection: striking the balance between precision and recall}. In \textit{Advances in Social Networks Analysis and Mining (ASONAM), 2016 IEEE/ACM International Conference on} (pp. 533–540). IEEE.

\phantomsection
\label{csl:61}Moturu, S. T., \& Liu, H. (2009). {Evaluating the trustworthiness of Wikipedia articles through quality and credibility}. In \textit{Proceedings of the 5th international symposium on wikis and open collaboration} (pp. 1–2).

\phantomsection
\label{csl:2}Nami Sumida, M. W., \& Mitchell, A. (2019). {The role of social media in news}. \textit{Pew Research Center - Journalism and Media}. Retrieved from \url{https://www.journalism.org/2019/04/23/the-role-of-social-media-in-news/}

\phantomsection
\label{csl:47}Nataraj, L., Mohammed, T. M., Manjunath, B. S., Chandrasekaran, S., Flenner, A., Bappy, J. H., \& Roy-Chowdhury, A. K. (2019). {Detecting GAN generated fake images using co-occurrence matrices}. \textit{arXiv preprint arXiv:1903.06836}.

\phantomsection
\label{csl:37}Nguyen, N. P., Yan, G., Thai, M. T., \& Eidenbenz, S. (2012). {Containment of misinformation spread in online social networks}. In \textit{Proceedings of the 4th Annual ACM Web Science Conference} (pp. 213–222).

\phantomsection
\label{csl:90}Nied, A. C., Stewart, L., Spiro, E., \& Starbird, K. (2017). {Alternative narratives of crisis events: Communities and social botnets engaged on social media}. In \textit{Companion of the 2017 ACM Conference on Computer Supported Cooperative Work and Social Computing} (pp. 263–266). ACM.

\phantomsection
\label{csl:46}Odena, A., Dumoulin, V., \& Olah, C. (2016). {Deconvolution and checkerboard artifacts}. \textit{Distill}, \textit{1}(10), e3.

\phantomsection
\label{csl:118}Pamungkas, E. W., Basile, V., \& Patti, V. (2019). {Stance classification for rumour analysis in Twitter: Exploiting affective information and conversation structure}. \textit{arXiv preprint arXiv:1901.01911}.

\phantomsection
\label{csl:38}Pariser, E. (2011). \textit{{The filter bubble: How the new personalized web is changing what we read and how we think}}. Penguin.

\phantomsection
\label{csl:86}Parkinson, H. J. (2016). {Click and elect: how fake news helped Donald Trump win a real election}. The Guardian. Retrieved from \url{https://goo.gl/DJiWNd}

\phantomsection
\label{csl:21}Parth M.N., S. B. (30AD). {Rumors of child-kidnapping gangs and other WhatsApp hoaxes are getting people killed in India}. \textit{Los Angeles Times}.

\phantomsection
\label{csl:22}Patrick J. McDonnell, C. S. (20AD). {When fake news kills: Lynchings in Mexico are linked to viral child-kidnap rumors}. \textit{Los Angeles Times}.

\phantomsection
\label{csl:78}Pennycook, G., \& Rand, D. (2019, January 19). {Why Do People Fall for Fake News?}. The New York Times. Retrieved from \url{https://www.nytimes.com/2019/01/19/opinion/sunday/fake-news.html}

\phantomsection
\label{csl:153}Pennycook, G., \& Rand, D. G. (2019). {Lazy, not biased: Susceptibility to partisan fake news is better explained by lack of reasoning than by motivated reasoning}. \textit{Cognition}, \textit{188}, 39–50. Elsevier.

\phantomsection
\label{csl:151}Pham, C. V., Phu, Q. V., Hoang, H. X., Pei, J., \& Thai, M. T. (2019). {Minimum budget for misinformation blocking in online social networks}. \textit{Journal of Combinatorial Optimization}, \textit{38}(4), 1101–1127. Springer.

\phantomsection
\label{csl:7}Pratiwi, I. Y. R., Asmara, R. A., \& Rahutomo, F. (2017). {Study of hoax news detection using naïve bayes classifier in Indonesian language}. In \textit{2017 11th International Conference on Information {\&} Communication Technology and System ({ICTS})}. {IEEE}. Retrieved from \url{https://doi.org/10.1109\%2Ficts.2017.8265649}

\phantomsection
\label{csl:114}Qian, F., Gong, C., Sharma, K., \& Liu, Y. (2018). {Neural User Response Generator: Fake News Detection with Collective User Intelligence.}. In \textit{IJCAI} (pp. 3834–3840).

\phantomsection
\label{csl:121}Radford, A., Narasimhan, K., Salimans, T., \& Sutskever, I. (2018). {Improving language understanding by generative pre-training}.

\phantomsection
\label{csl:122}Radford, A., Wu, J., Child, R., Luan, D., Amodei, D., \& Sutskever, I. (2019). {Language models are unsupervised multitask learners}. \textit{OpenAI Blog}, \textit{1}(8), 9.

\phantomsection
\label{csl:84}Ratkiewicz, J., Conover, M., Meiss, M. R., Gon{\c{c}}alves, B., Flammini, A., \& Menczer, F. (2011). {Detecting and tracking political abuse in social media.}. \textit{ICWSM}, \textit{11}, 297–304.

\phantomsection
\label{csl:103}Ratkiewicz, J., Conover, M., Meiss, M., Gon{\c{c}}alves, B., Patil, S., Flammini, A., \& Menczer, F. (2011). {Truthy: mapping the spread of astroturf in microblog streams}. In \textit{World Wide Web Companion} (pp. 249–252). ACM.

\phantomsection
\label{csl:13}Rosnow, R. L. (1991). {Inside rumor: A personal journey.}. \textit{American Psychologist}, \textit{46}(5), 484–496. American Psychological Association ({APA}). Retrieved from \url{https://doi.org/10.1037\%2F0003-066x.46.5.484}

\phantomsection
\label{csl:69}Rosnow, R. L., \& Fine, G. A. (1976). \textit{{Rumor and gossip: The social psychology of hearsay.}}. Elsevier.

\phantomsection
\label{csl:164}Salem, F. K. A., Al Feel, R., Elbassuoni, S., Jaber, M., \& Farah, M. (2019). {FA-KES: a fake news dataset around the Syrian war}. In \textit{Proceedings of the International AAAI Conference on Web and Social Media} (Vol. 13, pp. 573–582).

\phantomsection
\label{csl:8}Santoso, I., Yohansen, I., Nealson, Warnars, H. L. H. S., \& Hashimoto, K. (2017). {Early investigation of proposed hoax detection for decreasing hoax in social media}. In \textit{2017 {IEEE} International Conference on Cybernetics and Computational Intelligence ({CyberneticsCom})}. {IEEE}. Retrieved from \url{https://doi.org/10.1109\%2Fcyberneticscom.2017.8311705}

\phantomsection
\label{csl:55}Scheufele, D. A., \& Krause, N. M. (2019). {Science audiences, misinformation, and fake news}. \textit{Proceedings of the National Academy of Sciences}, \textit{116}(16), 7662–7669. National Academy of Sciences. Retrieved from \url{https://www.pnas.org/content/116/16/7662}

\phantomsection
\label{csl:28}Shao, C., Ciampaglia, G. L., Varol, O., Flammini, A., \& Menczer, F. (2017). {The spread of fake news by social bots}. \textit{arXiv preprint arXiv:1707.07592}, \textit{96}, 104. ArXiv e-prints.

\phantomsection
\label{csl:94}Shao, C., Ciampaglia, G. L., Varol, O., Yang, K.-C., Flammini, A., \& Menczer, F. (2018). {The spread of low-credibility content by social bots}. \textit{Nature Communications}, \textit{9}(1). Springer Science and Business Media {LLC}. Retrieved from \url{https://doi.org/10.1038\%2Fs41467-018-06930-7}

\phantomsection
\label{csl:93}Shao, C., Hui, P. M., Wang, L., Jiang, X., Flammini, A., Menczer, F., \& Ciampaglia, G. L. (2018). {Anatomy of an online misinformation network}. \textit{PLoS ONE}, \textit{13}(4), 1–23.

\phantomsection
\label{csl:29}Sharma, K., Qian, F., Jiang, H., Ruchansky, N., Zhang, M., \& Liu, Y. (2019). {Combating fake news: A survey on identification and mitigation techniques}. \textit{ACM Transactions on Intelligent Systems and Technology (TIST)}, \textit{10}(3), 1–42. ACM New York, NY, USA.

\phantomsection
\label{csl:1}Shearer, E., \& Matsa, K. E. (2018). {News Use Across Social Media Platforms 2018}. \textit{Pew Research Center - Journalism and Media}. Retrieved from \url{https://www.journalism.org/2018/09/10/news-use-across-social-media-platforms-2018/}

\phantomsection
\label{csl:126}Shi, B., \& Weninger, T. (2016). {Discriminative predicate path mining for fact checking in knowledge graphs}. \textit{Knowledge-based systems}, \textit{104}, 123–133. Elsevier.

\phantomsection
\label{csl:127}Shi, B., \& Weninger, T. (2016). {Fact checking in heterogeneous information networks}. In \textit{Proceedings of the 25th International Conference Companion on World Wide Web} (pp. 101–102).

\phantomsection
\label{csl:26}Shu, K., \& Liu, H. (2019). {Detecting fake news on social media}. \textit{Synthesis Lectures on Data Mining and Knowledge Discovery}, \textit{11}(3), 1–129. Morgan \& Claypool Publishers.

\phantomsection
\label{csl:147}Shu, K., Bernard, H. R., \& Liu, H. (2018). {Studying Fake News via Network Analysis: Detection and Mitigation}. In \textit{Lecture Notes in Social Networks} (pp. 43–65). Springer International Publishing. Retrieved from \url{https://doi.org/10.1007\%2F978-3-319-94105-9_3}

\phantomsection
\label{csl:166}Shu, K., Mahudeswaran, D., Wang, S., Lee, D., \& Liu, H. (2018). {FakeNewsNet: A Data Repository with News Content, Social Context and Dynamic Information for Studying Fake News on Social Media}. \textit{arXiv preprint arXiv:1809.01286}.

\phantomsection
\label{csl:24}Shu, K., Sliva, A., Wang, S., Tang, J., \& Liu, H. (2017). {Fake news detection on social media: A data mining perspective}. \textit{ACM SIGKDD Explorations Newsletter}, \textit{19}(1), 22–36. ACM New York, NY, USA.

\phantomsection
\label{csl:116}Shu, K., Zhou, X., Wang, S., Zafarani, R., \& Liu, H. (2019). {The role of user profiles for fake news detection}. In \textit{Proceedings of the 2019 IEEE/ACM International Conference on Advances in Social Networks Analysis and Mining} (pp. 436–439).

\phantomsection
\label{csl:57}Starbird, K. (2017). {Examining the alternative media ecosystem through the production of alternative narratives of mass shooting events on Twitter}. In \textit{Eleventh International AAAI Conference on Web and Social Media}.

\phantomsection
\label{csl:89}Starbird, K. (2017). {Examining the Alternative Media Ecosystem through the Production of Alternative Narratives of Mass Shooting Events on Twitter}. \textit{Icwsm (2017)}, (Icwsm), 230–239. Retrieved from \url{https://faculty.washington.edu/kstarbi/Alt{\_}Narratives{\_}ICWSM17-CameraReady.pdf{\%}0Ahttp://faculty.washington.edu/kstarbi/Alt{\_}Narratives{\_}ICWSM17-CameraReady.pdf{\%}0Ahttps://aaai.org/ocs/index.php/ICWSM/ICWSM17/paper/view/15603}

\phantomsection
\label{csl:95}Stella, M., Ferrara, E., \& {De Domenico}, M. (2018). {Bots sustain and inflate striking opposition in online social systems}, 1–10. Retrieved from \url{http://arxiv.org/abs/1802.07292}

\phantomsection
\label{csl:15}Sunstein, C. R., \& Vermeule, A. (2009). {Conspiracy theories: Causes and cures}. \textit{Journal of Political Philosophy}, \textit{17}(2), 202–227. Wiley Online Library.

\phantomsection
\label{csl:67}Swire, B., Ecker, {U. K. H., \& Lewandowsky, S. (2017). {The Role of Familiarity in Correcting Inaccurate Information}. \textit{JOURNAL OF EXPERIMENTAL PSYCHOLOGY-LEARNING MEMORY AND COGNITION}, \textit{43}(12), 1948–1961. AMER PSYCHOLOGICAL ASSOC}.

\phantomsection
\label{csl:167}Tacchini, E., Ballarin, G., Della Vedova, M. L., Moret, S., \& Alfaro, L. de. (2017). {Some like it hoax: Automated fake news detection in social networks}. \textit{arXiv preprint arXiv:1704.07506}.

\phantomsection
\label{csl:157}Tandoc Jr, E. C. (2019). {The facts of fake news: A research review}. \textit{Sociology Compass}, \textit{13}(9), e12724. Wiley Online Library.

\phantomsection
\label{csl:40}Tariq, S., Lee, S., Kim, H., Shin, Y., \& Woo, S. S. (2018). {Detecting both machine and human created fake face images in the wild}. In \textit{Proceedings of the 2nd International Workshop on Multimedia Privacy and Security} (pp. 81–87). ACM.

\phantomsection
\label{csl:16}Taylor, A. (2016). {Before ‘fake news,’ there was Soviet ‘disinformation’}. \textit{The Washington Post}. \url{https://www.washingtonpost.com/news/worldviews/wp/2016/11/26/before-fake-news-there-was-soviet-disinformation/.} Retrieved from \url{https://www.washingtonpost.com/news/worldviews/wp/2016/11/26/before-fake-news-there-was-soviet-disinformation/}

\phantomsection
\label{csl:104}Thomas, K., Grier, C., \& Paxson, V. (2012). {Adapting social spam infrastructure for political censorship}. In \textit{Conference on Large-Scale Exploits and Emergent Threats}. USENIX.

\phantomsection
\label{csl:83}Timberg, C., \& Dwoskin, E. (2018, July). {Twitter is Sweeping out Fake Accounts Like Never Before, Putting User Growth at Risk}. Retrieved from \url{https://www.washingtonpost.com/technology/2018/07/06/twitter-is-sweeping-out-fake-accounts-like-never-before-putting-user-growth-risk/}

\phantomsection
\label{csl:141}Timsit, A. (12AD). {In the age of fake news, here’s how schools are teaching kids to think like fact-checkers}. \textit{Quartz}. Retrieved from \url{https://qz.com/1533747/in-the-age-of-fake-news-heres-how-schools-are-teaching-kids-to-think-like-fact-checkers/}

\phantomsection
\label{csl:150}Tong, A., Du, D.-Z., \& Wu, W. (2018). {On Misinformation Containment in Online Social Networks}. In S. Bengio, H. Wallach, H. Larochelle, K. Grauman, N. Cesa-Bianchi, \& R. Garnett (Eds.), \textit{Advances in Neural Information Processing Systems 31} (pp. 341–351). Curran Associates, Inc. Retrieved from \url{http://papers.nips.cc/paper/7317-on-misinformation-containment-in-online-social-networks.pdf}

\phantomsection
\label{csl:113}Tschiatschek, S., Singla, A., Gomez Rodriguez, M., Merchant, A., \& Krause, A. (2018). {Fake news detection in social networks via crowd signals}. In \textit{Companion Proceedings of the The Web Conference 2018} (pp. 517–524). International World Wide Web Conferences Steering Committee.

\phantomsection
\label{csl:142}Tugend, A. (20AD). {These Students Are Learning About Fake News and How to Spot It}. \textit{The New York Times}. Retrieved from \url{https://www.nytimes.com/2020/02/20/education/learning/news-literacy-2016-election.html}

\phantomsection
\label{csl:146}Van Duyn, E., \& Collier, J. (2019). {Priming and fake news: The effects of elite discourse on evaluations of news media}. \textit{Mass Communication and Society}, \textit{22}(1), 29–48. Taylor \& Francis.

\phantomsection
\label{csl:82}Varol, O., Ferrara, E., Davis, C. A., Menczer, F., \& Flammini, A. (2017). {Online human-bot interactions: Detection, estimation, and characterization}. In \textit{ICWSM} (pp. 280–289).

\phantomsection
\label{csl:9}Vedova, M. L. D., Tacchini, E., Moret, S., Ballarin, G., DiPierro, M., \& Alfaro, L. de. (2018). {Automatic Online Fake News Detection Combining Content and Social Signals}. In \textit{2018 22nd Conference of Open Innovations Association ({FRUCT})}. {IEEE}. Retrieved from \url{https://doi.org/10.23919\%2Ffruct.2018.8468301}

\phantomsection
\label{csl:117}Vo, N., \& Lee, K. (2018). {The rise of guardians: Fact-checking url recommendation to combat fake news}. In \textit{The 41st International ACM SIGIR Conference on Research \& Development in Information Retrieval} (pp. 275–284). ACM.

\phantomsection
\label{csl:27}Vosoughi, S., Roy, D., \& Aral, S. (2018). {The spread of true and false news online}. \textit{Science}, \textit{359}(6380), 1146–1151. American Association for the Advancement of Science.

\phantomsection
\label{csl:11}Vukovi{\'{c}}, M., Pripu{\v{z}}i{\'{c}}, K., \& Belani, H. (2009). {An Intelligent Automatic Hoax Detection System}. In \textit{Knowledge-Based and Intelligent Information and Engineering Systems} (pp. 318–325). Springer Berlin Heidelberg. Retrieved from \url{https://doi.org/10.1007\%2F978-3-642-04595-0_39}

\phantomsection
\label{csl:72}Waddington, K. (2012). \textit{{Gossip and organizations}}. Routledge.

\phantomsection
\label{csl:102}Wang, G., Mohanlal, M., Wilson, C., Wang, X., Metzger, M., Zheng, H., \& Zhao, B. Y. (2013). {Social Turing Tests: Crowdsourcing Sybil Detection}. \textit{arXiv preprint arXiv:1205.3856}. Internet Society.

\phantomsection
\label{csl:162}Wang, W. Y. (2017). { liar, liar pants on fire: A new benchmark dataset for fake news detection}. \textit{arXiv preprint arXiv:1705.00648}.

\phantomsection
\label{csl:52}Wang, Y., Ma, F., Jin, Z., Yuan, Y., Xun, G., Jha, K., Su, L., et al. (2018). {Eann: Event adversarial neural networks for multi-modal fake news detection}. In \textit{Proceedings of the 24th acm sigkdd international conference on knowledge discovery \& data mining} (pp. 849–857).

\phantomsection
\label{csl:132}Wu, K., Yang, S., \& Zhu, K. Q. (2015). {False rumors detection on sina weibo by propagation structures}. In \textit{2015 IEEE 31st international conference on data engineering} (pp. 651–662). IEEE.

\phantomsection
\label{csl:3}Wu, L., Morstatter, F., Carley, K. M., \& Liu, H. (2019). {Misinformation in Social Media}. \textit{{ACM} {SIGKDD} Explorations Newsletter}, \textit{21}(2), 80–90. Association for Computing Machinery ({ACM}). Retrieved from \url{https://doi.org/10.1145\%2F3373464.3373475}

\phantomsection
\label{csl:109}Xie, Y., Yu, F., Achan, K., Panigrahy, R., Hulten, G., \& Osipkov, I. (2008). {Spamming botnets: signatures and characteristics}. \textit{ACM SIGCOMM Computer Communication Review}, \textit{38}(4), 171–182. ACM.

\phantomsection
\label{csl:25}Yang, S., Shu, K., Wang, S., Gu, R., Wu, F., \& Liu, H. (2019). {Unsupervised Fake News Detection on Social Media: A Generative Approach}. \textit{Proceedings of the {AAAI} Conference on Artificial Intelligence}, \textit{33}, 5644–5651. Association for the Advancement of Artificial Intelligence ({AAAI}). Retrieved from \url{https://doi.org/10.1609\%2Faaai.v33i01.33015644}

\phantomsection
\label{csl:51}Yang, X., Li, Y., \& Lyu, S. (2019). {Exposing deep fakes using inconsistent head poses}. In \textit{ICASSP 2019-2019 IEEE International Conference on Acoustics, Speech and Signal Processing (ICASSP)} (pp. 8261–8265). IEEE.

\phantomsection
\label{csl:53}Yang, Y., Zheng, L., Zhang, J., Cui, Q., Li, Z., \& Yu, P. S. (2018). {TI-CNN: Convolutional neural networks for fake news detection}. \textit{arXiv preprint arXiv:1806.00749}.

\phantomsection
\label{csl:129}Yin, X., Han, J., \& Philip, S. Y. (2008). {Truth discovery with multiple conflicting information providers on the web}. \textit{IEEE Transactions on Knowledge and Data Engineering}, \textit{20}(6), 796–808. IEEE.

\phantomsection
\label{csl:36}Zafarani, R., Abbasi, M. A., \& Liu, H. (n.d.). {Information Diffusion in Social Media}. In \textit{Social Media Mining} (pp. 179–214). Cambridge University Press. Retrieved from \url{https://doi.org/10.1017\%2Fcbo9781139088510.008}

\phantomsection
\label{csl:119}Zellers, R., Holtzman, A., Rashkin, H., Bisk, Y., Farhadi, A., Roesner, F., \& Choi, Y. (2019). {Defending against neural fake news}. In \textit{Advances in Neural Information Processing Systems} (pp. 9051–9062).

\phantomsection
\label{csl:110}Zhang, C. M., \& Paxson, V. (2011). {Detecting and Analyzing Automated Activity on Twitter. In N. Spring and G. Riley (Eds.)}. \textit{Passive and Active Measurement. PAM 2011}, \textit{LNCS 6579}, 102–111.

\phantomsection
\label{csl:130}Zhang, D. Y., Wang, D., \& Zhang, Y. (2017). {Constraint-aware dynamic truth discovery in big data social media sensing}. In \textit{2017 IEEE International Conference on Big Data (Big Data)} (pp. 57–66). IEEE.

\phantomsection
\label{csl:45}Zhang, X., Karaman, S., \& Chang, S.-F. (2019). {Detecting and simulating artifacts in gan fake images}. \textit{arXiv preprint arXiv:1907.06515}.

\phantomsection
\label{csl:43}Zhu, J.-Y., Park, T., Isola, P., \& Efros, A. A. (2017). {Unpaired image-to-image translation using cycle-consistent adversarial networks}. In \textit{Proceedings of the IEEE international conference on computer vision} (pp. 2223–2232).

\phantomsection
\label{csl:163}Zubiaga, A. (2018). {Learning class-specific word representations for early detection of hoaxes in social media}. \textit{arXiv preprint arXiv:1801.07311}.

\phantomsection
\label{csl:160}Zubiaga, A., Liakata, M., Procter, R., Hoi, G. W. S., \& Tolmie, P. (2016). {Analysing how people orient to and spread rumours in social media by looking at conversational threads}. \textit{PloS one}, \textit{11}(3). Public Library of Science.

\phantomsection
\label{csl:14}Tempel, J. van der, \& Alcock, J. E. (2015). {Relationships between conspiracy mentality hyperactive agency detection, and schizotypy: Supernatural forces at work?}. \textit{Personality and Individual Differences}, \textit{82}, 136–141. Elsevier {BV}. Retrieved from \url{https://doi.org/10.1016\%2Fj.paid.2015.03.010}

\phantomsection
\label{csl:18}{Department of State.}. (1981). {Forgery, Disinformation and Political Operation.}. \textit{Department of State Bulletin}, \textit{81}(2056), 52–5.

\end{document}